\documentclass[onecolumn,preprint,english,aps,showpacs,superscriptaddress,notitlepage]{revtex4-2}
\usepackage{amsmath,amssymb}
\usepackage{graphicx}
\usepackage{txfonts}
\usepackage{color}
\usepackage{hyperref}
\usepackage{bbold}
\usepackage{framed}
\usepackage{wrapfig}
\usepackage{multirow}
\usepackage{ulem}
\usepackage[letterpaper]{geometry}
\global\long\def\av#1{\left\langle #1 \right\rangle }

\global\long\def\up{\uparrow}
\global\long\def\down{\downarrow}

\renewcommand{\vec}[1]{\boldsymbol{#1}}
\newcommand{\beq}{\begin{equation}}
\newcommand{\eeq}{\end{equation}}
\newcommand{\bea}{\begin{eqnarray}}
\newcommand{\eea}{\end{eqnarray}}

\begin{document}

\title{ Cascade of transitions in twisted and non-twisted graphene layers within the van Hove scenario }

\author{Dmitry V. Chichinadze}
\affiliation{School of Physics and Astronomy,
University of Minnesota, Minneapolis, MN 55455, USA}
\email{chich013@umn.edu}
\author{Laura Classen}
\affiliation{Max Planck Institute for Solid State Research, D-70569 Stuttgart, Germany}
\author{Yuxuan Wang}
\affiliation{Department of Physics, University of Florida, Gainesville, Florida 32601}
\author{Andrey V. Chubukov}
\affiliation{School of Physics and Astronomy,
University of Minnesota, Minneapolis, MN 55455, USA}
\affiliation{W. I. Fine Theoretical Physics Institute, University of Minnesota, Minneapolis, Minnesota 55455, USA}
\email{achubuko@umn.edu}
\begin{abstract}
\bf{
Motivated by measurements of compressibility and STM spectra in twisted bilayer graphene, we analyze the pattern of symmetry breaking for itinerant fermions near a van Hove singularity. Making use of an approximate SU(4) symmetry of the Landau functional, we show that the structure of the spin/isospin order parameter changes with increasing filling via a cascade of transitions. We compute the feedback from different spin/isospin orders on fermions and argue that each order splits the initially 4-fold degenerate van Hove peak in a particular fashion, consistent with the STM data and compressibility measurements, providing a unified interpretation of the cascade of transitions in twisted bilayer graphene. Our results follow from a generic analysis of an SU(4)-symmetric Landau functional and are valid beyond a specific underlying fermionic model. We argue that an analogous van Hove scenario explains the cascade of phase transitions in non-twisted Bernal bilayer and rhombohedral trilayer graphene. }
   \end{abstract}

\maketitle

\section*{\uppercase{Introduction}}

\begin{figure*}[h!]
\center{\includegraphics[width=\textwidth]{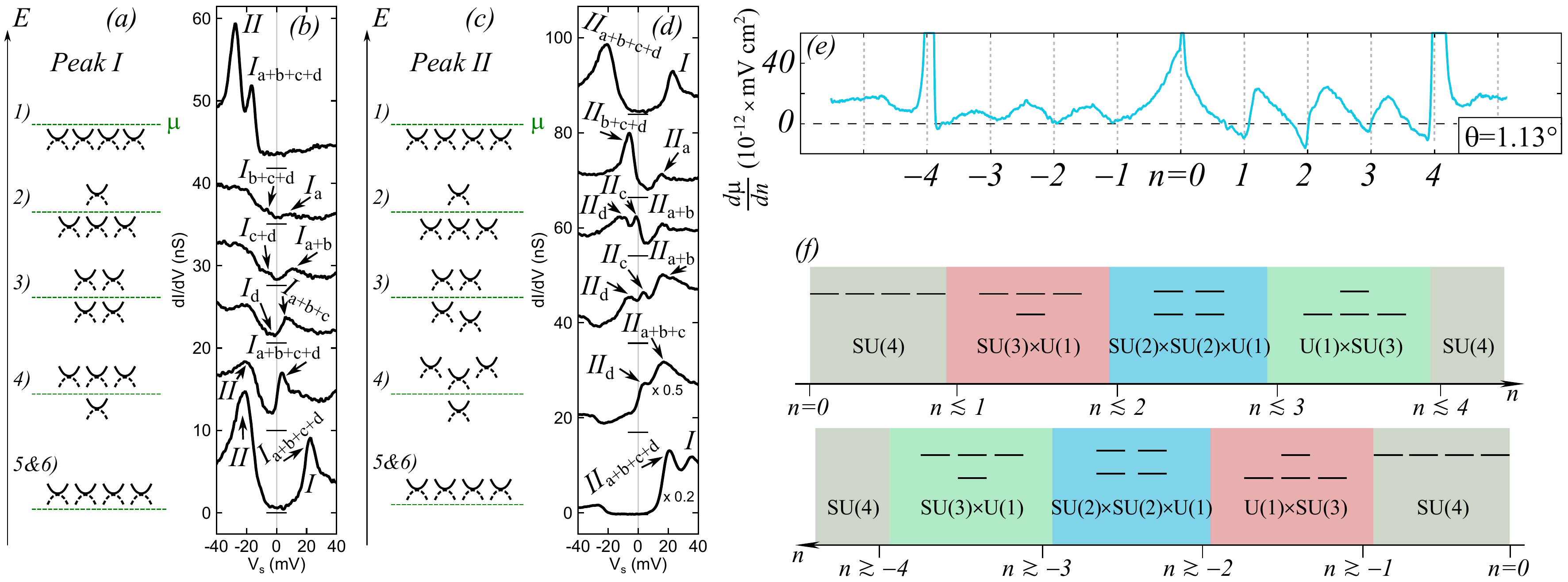}}
\centering{}\caption{ \textbf{The cascade of electronic transitions in twisted bilayer graphene.}
 (a)--(d):
 the proposed splitting  of the initially 4-fold (spin and valley) degenerate vH
 peak upon
 raising the electron
 filling $n$ 
 (panels  (a), (b)) 
  and hole filling
  (panels   (c), (d)),
   %
  and the corresponding STM data for twist angle of $1.06^{\circ}$, reproduced
  with permission from the authors of Ref. \cite{ali_2}.
  The van Hove  peaks in the conduction (valence) bands are labeled by $I$ $(II)$, and
  subscripts
   a,b,c,d label the  4 peak components.
 (e):
  experimental data for inverse compressibility with seesaw features, interpreted as a cascade of phase transitions.
  Reproduced with permission from the authors of  Ref. \cite{Zondiner2020}. The data are for twist angle $\theta = 1.13^{\circ}$,
 (f)
 The
schematic phase diagram of TBG upon electron or hole doping, based on comparison between the theory and the STM data.
 Within our model, we obtained SU(3) $\times$ U(1)  (U(1)$\times$ SU(3)) symmetry
 for both electron and hole doping at
 $1< |n| <2$  ($3<|n|<4$), corresponding to 3-1 (1-3) splitting, but it is
  reduced to  SU(2) $\times$ U(1) $\times$ U(1) (U(1)$\times$ U(1) $\times$ SU(2))
 if there is an intermediate phase with 2-1-1 (1-1-2) splitting, as STM data for hole doping likely indicate.}

 \label{cascade_pic}
\end{figure*}

Twisted bilayer graphene (TBG) is a two-dimensional correlated electronic system, which
exhibits superconductivity \cite{Cao2018SC,Yankowitz2019tuning,Cao2020NematicSC} and correlated phases
 \cite{Cao2018insulator,Xie2019,Jiang2019nematic,Chen2020,Saito2021,Xie2021FCI,
 Sharpe2019,Serlin2020AQHE,Sharpe2021,Tschirhart2021}.
 The focus of our work is the analysis of a cascade of phase transitions near integer fillings $|n|= 1,2,3,4$,
 detected in
   STM and electronic compressibility  measurements
  \cite{Xie2019,ali_2,Zondiner2020} (panels
  (a)-(e)
   in Fig. \ref{cascade_pic}).   Compressibility measurements
  show sharp seesaw features of $d\mu/dn$ near integer $|n|$,  and STM data show that around
   each of these  $n$ a peak in the
   density of states splits, and one of its  components appears on the other side of the Fermi level.
  For the interpretation, the authors of~\cite{ali_2} adopted a strong coupling approach and associated  the observed STM peaks with narrow sub-bands. They argued that at each transition one sub-band crosses the Fermi level, moves
    away from it, and becomes
    incoherent.  The authors of \cite{Zondiner2020} interpreted compressibility
     data within a moderate coupling scenario of a 4-fold spin/isospin degenerate band and argued
   that the cascade  can be understood as a
  series
 of interaction-driven transitions. They conjectured that at, e.g.,  electronic doping
   one of the bands gets completely filled  at each transition,
   while  the occupation of the remaining ones gets
  depleted; mirror symmetric behavior holds for hole doping.

 In this communication we propose the
  scenario in which
  the cascade of transitions is  caused by the development of particle-hole orders, like in
    \cite{Zondiner2020}, but we specifically identify the STM peaks with van Hove (vH) singularities.  We
argue that the components of the initially 4-fold degenerate vH peak move through the Fermi level one by one, but remain close to it.  The split peaks recombine into a single 4-fold peak at $|n| \lesssim 4$, when
 electronic order vanishes.
 Our scenario is 
   illustrated
   in panels
  (a), (c), and (f)
  in Fig. \ref{cascade_pic}.

A  cascade of transitions has been observed near van Hove doping
  in less correlated non-twisted Bernal bilayer (BBG) and rhombohedral trilayer graphene (RTG) ~\cite{BBG_exp2022,Seiler2022,Zhou2021SC,Zhou2021cascade}. We show that our vH scenario equally explains the  sequence of transitions in these materials.
  We believe that the similarity
   between the ordered states and electronic reconstruction  in BBG/RTG and in TBG
  supports a
 moderate coupling vH-based approach.  We emphasize, however, that we use this  approach specifically to describe the cascade of phase transitions with doping. A strong coupling approach is
 needed for
 explaining the
 insulating behavior of TBG near integer fillings.

We further emphasize that  (i) vH peaks  have been observed  in TBG at different twisting  angles \cite{Li2010,Jiang2019nematic}, (ii) are  present in
    the electronic dispersion, obtained in first-principle calculations, and in the one
    renormalized by the interaction, even if the bottom of the dispersion moves away from  Dirac points \cite{Kang2021PRL}, and (iii) the cascade of transitions, observed in
   magic-angle twisted trilayer graphene, has been argued to be triggered by vH peaks, at least at high displacement fields~\cite{Park2021TTG}. The vH scenario has been also discussed in context of chiral density wave and superconductivity in TBG (see e.g., \cite{PhysRevLett.121.217001,PhysRevX.8.041041,PhysRevLett.122.026801,PhysRevB.98.205151,ren2020spectroscopic}).

The summary of our results
  for TBG
 is presented in Fig. \ref{cascade_pic} along with the experimental data from Refs.
\cite{ali_2,Zondiner2020}. We label the vH peaks in conduction (valence) bands in  by $I$ ($II$) and label
peak components by $\text{a,b,c,d}$.
 Our interpretation of the STM data from Ref. \cite{ali_2}
      for electron doping
      (panel (b))
       is the following:
      as
        the system moves away from charge neutrality,
        the 4-fold degenerate peak $I_{\mathrm{a+b+c+d}}$ approaches $\mu$
         from above,
         and at $n  \leq 1$,
           splits  in a 3-1 fashion: three components $I_{\mathrm{a+b+c}}$
           stay above $\mu$, and one component, $I_{\mathrm{d}}$,
             jumps to
             below the Fermi level, but remains close to it.
        At
        $n  \leq 2$, the three components again come close to  $\mu$,
         and  the vH peak $I_{\mathrm{a+b+c}}$
         splits in 2-2 fashion,
          such that
         $I_{\mathrm{a+b}}$
         moves back, while $I_{\mathrm{c}}$ jumps to below the Fermi level and merges with $I_{\mathrm{d}}$ into $I_{\mathrm{c+d}}$. At $n \lesssim 3$, $I_{\mathrm{a+b}}$ splits and $I_{\mathrm{b}}$ jumps to below the Fermi level and merges with $I_{\mathrm{c+d}}$ into $I_{\mathrm{b+c+d}}$. Finally, at $n \lesssim 4$, the last component $I_{\mathrm{a}}$ jumps
         across the  Fermi level and merges with three other components into 4-fold degenerate $I_{\mathrm{a+b+c+d}}$.  For hole doping
          the overall evolution is the same, but
          the data seem to show a more gradual behavior:
         the components of peak $II$ in
         panel (d)
         cross the Fermi level one-by-one,
         indicating the presence of an  intermediate state between $3-1$ and $1-3$ ones.

 We
  consider
  these data as evidence that once the 4-fold degenerate vH peak gets close to $\mu$ at $|n| \sim 1$ (peak $I$ for $n >0$ and peak $II$ for $n <0$), the system
    develops a vH-induced particle-hole order. The order exists between
     $|n| \lesssim 1$ and $|n| \lesssim 4$
     and reconstructs the fermionic spectra, pushing vH peaks in some bands
     above $\mu$
    and in other band(s)
    below $\mu$.  The structure of the order changes near $|n|=2$ and $|n|=3$,
    via  first-order
     transitions,
     and this changes
      the splitting of the vH peak and simultaneously gives rise to sharp changes in the
    compressibility
    (see Fig. \ref{cascade_pic} (e)).

 Here, we present the theoretical description of this scenario within the model of interacting
    itinerant electrons whose band structure has a vH singularity near $\mu$.  We use as
      an input
      our earlier result~\cite{Chichinadze2020magnet,Chichinadze2021su4}  that the increased density of states near the vH singularity enables a spontaneous symmetry breaking in spin and valley spaces
      (for particle-hole orders with zero transferred momentum electronic DOS has to exceed a threshold
       in order to satisfy the Stoner-type criterion).
     For a model with intra-site (Hubbard) and assisted hopping
      interactions within a given hexagon
     \cite{Kang2019PRL}, we found
     15 particle-hole order parameters, for which the couplings are  attractive, near-equal,
    and larger than  for other order parameters.
    They describe intra-valley order at zero momentum ($\mathbf{Q}=0$) and inter-valley density waves ($\mathbf{Q}\neq 0$).
   These 15 order parameters are described by  $4\times4$  matrices in spin and valley spaces, specified by spin
   $\sigma$ and valley isospin $\tau$ and form  the adjoint representation of the SU(4) group. For shortness,  we
   call an order in $(\sigma, \tau)$ space a spin/isospin order.
   An SU(4)-symmetric order parameter manifold
    and the interplay between spin and isospin orders
     have been recently discussed in~\cite{PhysRevB.103.024506,PhysRevX.10.031034,PhysRevLett.121.087001,PhysRevB.98.245103,Kang2019PRL,PhysRevB.99.195120,PhysRevB.100.205131,Uchoa2019} -- these works provide
     additional motivation for us.
     Another input for our analysis are band structure
   calculations~\cite{Cea2019,PhysRevB.100.205114},
   which  reported the pinning of the vH singularity  to the chemical potential over the
   range of $n$.

   We derive and analyze the Landau free energy for an SU(4)-symmetric fermionic model.
   We argue that there are
   three sets of ordered states, which split the 4-fold degenerate vH peak
     in
     three different ways.
   The  first vH-induced instability splits the
   vH peak in a 3-1 fashion, with peaks for 3 degenerate bands
  shifting towards charge neutrality, and the remaining peak moving to below $\mu$.  As the magnitude of a
  spin/isospin order increases, the system  undergoes  a transition into a different ordered state, which
  splits the vH peak into
  a 2-2 fashion.
  The transition is
  first-order in our model, but in reality may
  occur via an intermediate phase with 2-1-1 splitting.
   At $|n| \sim 3$ the magnitude of
  the order starts decreasing as some of vH peak components move further away from $\mu$, and the system behavior
  goes in reverse - first the system undergoes a transition
      into an ordered state which gives rise to  1-3 vH peak splitting, again either via a direct first-order
      transition, or via an intermediate  phase with 1-1-2 splitting, and then, at even larger $|n| \leq 4$, the
      order vanishes, and all 4 vH peak components merge into a single vH peak below $\mu$.

We use the same approach for BBG and RTG.  The bands structures and Fermi surfaces of BBG and RTG are very similar, and we model both systems by an effective patch model of fermions, located in the vicinity of $\mathbf{K}$ and $\mathbf{K'}$ points in the BZ.  We find that the 15 leading instabilities are
   analogous to
   TBG: towards valley polarization and intra-valley spin order (both with $\textbf{Q}=0$) and towards inter-valley charge and spin density wave orders with  $\textbf{Q} = \textbf{K}-\textbf{K'}$ .
    We find that for
    a Hubbard interaction, these orders are described by the same SU(4)-symmetric Landau free energy functional,
    Eqs. (1) and (2),  as in TBG.
     Like in TBG, the first vH-induced transition is into a state with valley polarization and ferromagnetism in a single valley.
This order gives rise to $3-1$ splitting, which in the case of BBG/RTG gives rise to one larger and three smaller Fermi pockets. This splitting  is analogous to the one observed in the IF$_1$ state in the notations of Ref. \cite{BBG_exp2022}.
  The subsequent transition upon doping is into a state with  either pure valley charge order or ferromagnetic order in both valleys.  This state gives rise to $2-2$ splitting, which in BBG/RTG  gives rise to
    two larger and two smaller Fermi pockets. This is analogous to PIP$_2$ state \cite{BBG_exp2022}.  A
     potential intermediate state with $2-1-1$ splitting is analogous to PIP$_1$ state in \cite{BBG_exp2022}.

     \section*{\uppercase{Results}}

{\bf Cascade of transitions in TBG} ~~~
  Band structure calculations show that there are eight  bands within the flat-band regime of TBG, accounting
  for two spin projections, two valley degrees of freedom from the original graphene layers, and two sublattices
  of  the moir\'e superlattice.
  Four bands are with upward and four with downward dispersion, merging at Dirac points $\textbf{K}$ and $\textbf{K'}$.
  Upon electron
  (hole) doping the chemical potential moves up (down), simultaneously changing the filling of four bands.
  Each band displays a vH singularity.
  The vH singularities for four conduction (four valence) bands are at the same energy.
   It was argued that strong coupling renormalizations
    may
    shift the
  minimum of electron band to the $\Gamma$ point \cite{Cea2019,Kang2021PRL},
    but vH singularities remain even for the renormalized dispersion \cite{Guinea2018PNAS,Cea2019}.

We study the cascade  of phase transitions by analyzing the Landau free energy for the ordered phases of fermions
 with vH singularity near $\mu$. The order parameters are expectation values of fermionic bilinears, and the
 free energy can be obtained by
  departing from a microscopic model of vH fermions with 4-fermion Hubbard and
  assisted hopping
   interactions~\cite{Kang2019PRL}
  and
 integrating out fermions after performing a Hubbard-Stratonovich transformation.
Alternatively, one can write down the Landau free energy solely based on symmetries and fix parameters
 phenomenologically through comparison with experiments.
 In an earlier study~\cite{Chichinadze2021su4} we found that out of
  a large number of possible fermionic bilinears
  (143  in the 6-patch vH model and even larger number in 12-patch model)
 there are 15, for which the couplings are attractive and the largest by magnitude.  The set of 15  is composed of
 two subsets of 7 and 8 bilinears with a single coupling within each subset.
   7 bilinears  with coupling $\lambda_7$ are intra-valley with transferred momentum $\textbf{Q}=0$, and 8 with coupling $\lambda_8$  are inter-valley with a finite $\textbf{Q}$
  (Kekule-type states considered in \cite{PhysRevLett.128.156401}).

   The couplings $\lambda_7$ and $\lambda_8$ are not identical, but are close to each other~\cite{Chichinadze2020magnet}.
 In our analysis we treat $\lambda_7$ and $\lambda_8$ as equal, in which case the 15 bilinears form an adjoint representation of SU(4).  We checked
  that the cascade of transitions and the sequence of vH peak splitting is the same in the model with only
  7 bilinears (the case $\lambda_7 > \lambda_8$).  In the
    model with 8 bilinears
    there is a single ordered phase and no cascade.

For the SU(4) case, the
Landau
free energy up to fourth order is~\cite{Chichinadze2021su4} \footnote{Note, that the expression here uses a slightly different definition of prefactors in the free energy than the one in \cite{Chichinadze2021su4}.}
\begin{equation}
\begin{gathered}
\mathcal{F} = - \frac{\alpha}{2} \mathrm{Tr}[\hat{\Phi}^2] + \frac{\gamma}{3} \mathrm{Tr}[\hat{\Phi}^3] +
\frac{\beta}{4} \mathrm{Tr}[\hat{\Phi}^4] +\frac{\beta'}{4} \mathrm{Tr}[\hat{\Phi}^2]^2,
\end{gathered}
\label{Lagr_main}
\end{equation}
where $\hat{\Phi} = \sum_{j=1}^{15} \phi_j T^j$, $T^j$  are generators of SU(4),
  and  $\phi_j \sim
  f^{\dagger} T^j f$
 are fermionic bilinears,
 which we treat as Hubbard-Stratonovich fields ($f$ and $f^{(\dagger)}$ are
operators of electrons  near vH points).
 The term
  $\beta'$ does not appear within Hubbard-Stratonovich but is allowed by symmetry, and we keep it
  for generality.

   By construction, $\hat{\Phi}$ can be represented by a traceless matrix \cite{Hamermesh}.
  In the diagonal basis
\begin{equation}
\hat{\Phi} = \mathrm{diag}(\lambda_1, \lambda_2,\lambda_{3}, -(\lambda_1+\lambda_2+\lambda_{3})),
\label{phi_matrix}
\end{equation}
and the free energy is
\begin{align}
\mathcal{F} &= - \frac{\alpha}{2} \left( \sum_j^{3} \lambda_j^2 + (\sum_j^{3} \lambda_j)^2 \right) + \frac{\gamma}{3}
\left( \sum_j^{3} \lambda_j^3 - (\sum_j^{3} \lambda_j)^3 \right)  \notag\\
&+ \frac{\beta}{4} \left( \sum_j^{3} \lambda_j^4 + (\sum_j^{3} \lambda_j)^4 \right)
+ \frac{\beta'}{4}\left( \sum_j^{3} \lambda_j^2 + (\sum_j^{3} \lambda_j)^2 \right)^2.
\label{Lagr2_main}
\end{align}
 At $\gamma =0$, the order develops continuously when $\alpha$ changes sign and becomes positive.
At a finite $\gamma$, the transition  is necessarily first order and occurs already when $\alpha$ is negative.
 Below we restrict to $\alpha >0$, when the order is already finite
  and also set $\beta >0$, consistent with the Hubbard-Stratonovich analysis and the calculation of $\alpha,\beta$ for the tight-binding model near a vH singularity~\cite{Chichinadze2021su4}. We discuss the behavior of $\gamma$ below and in Supplementary Discussion V.
Minimizing $\mathcal{F}$  with respect to $\lambda_j$ ($j=1,2,3$), we find three solutions
(up to permutations of
$\lambda_j$): (i) $\lambda_1 = \lambda_2 = \lambda_3$; (ii) $\lambda_1 = \lambda_2 = -\lambda_3$; (iii)  $\lambda_1 = \lambda_2  \neq \lambda_3$ (see Supplementary
Discussion I for details).
 For the first solution, $\hat\Phi = \mathrm{diag} (\lambda, \lambda, \lambda, -3\lambda)$,
 and the broken symmetry is described by the coset  SU(4)/[SU(3)$\times$U(1)], where  SU(3) corresponds to the
 transformation within the subset of
 the
  first three components of  $\hat{\Phi}$,
 and U(1) to a rotation of the last component relative to the other three.
 The  ordered states in terms of expectation values of fermionic bilinears $\langle \phi_i \rangle$  are
  mixtures
   of spin/isospin order with particular ratios of spin and isospin components~\cite{Chichinadze2021su4}.
   For example,
   a pure intra-valley order is a part of this set, but a pure inter-valley order is not.
   The order parameter manifold has $15-8-1=6$ Goldstone modes.  The feedback of this order on fermions is 3-1 or
   1-3 splitting of vH peaks, depending on the sign of $\gamma$.
 For the second solution, $\hat\Phi = \mathrm{diag} (\lambda, \lambda, -\lambda, -\lambda)$, and
 the  broken symmetry
  is SU(4)/[SU(2)$\times$SU(2)$\times$U(1)], where the two SU(2)'s correspond to rotations within the subsets of the first two and the last two components of $\hat\Phi$,
 and U(1) corresponds to a rotation of one subset relative to the other.
  The  ordered states in terms of  $\langle \phi_i \rangle$
    include pure spin and isospin orders, e.g. intra-valley ferromagnetism
     and
  valley polarization, and various inter-valley density waves
  ~\cite{Chichinadze2021su4}. This manifold has  $15-6-1=8$ Goldstone modes.
     The feedback from such order on fermions leads to  2-2 splitting of the vH peaks.  Finally, the third solution
     describes a mixed state with $\hat\Phi = \mathrm{diag} (\lambda, \lambda, -\lambda_3, -2\lambda+ \lambda_3)$ and
 broken symmetry SU(4)/[SU(2)$\times$U(1)$\times$U(1)].
 The order parameter manifold contains $15-3-1-1=10$ Goldstone modes. The feedback on fermions leads to
 2-1-1 or 1-1-2 splitting of vH peaks.

\begin{figure}[h!]
\center{\includegraphics[width=\textwidth]{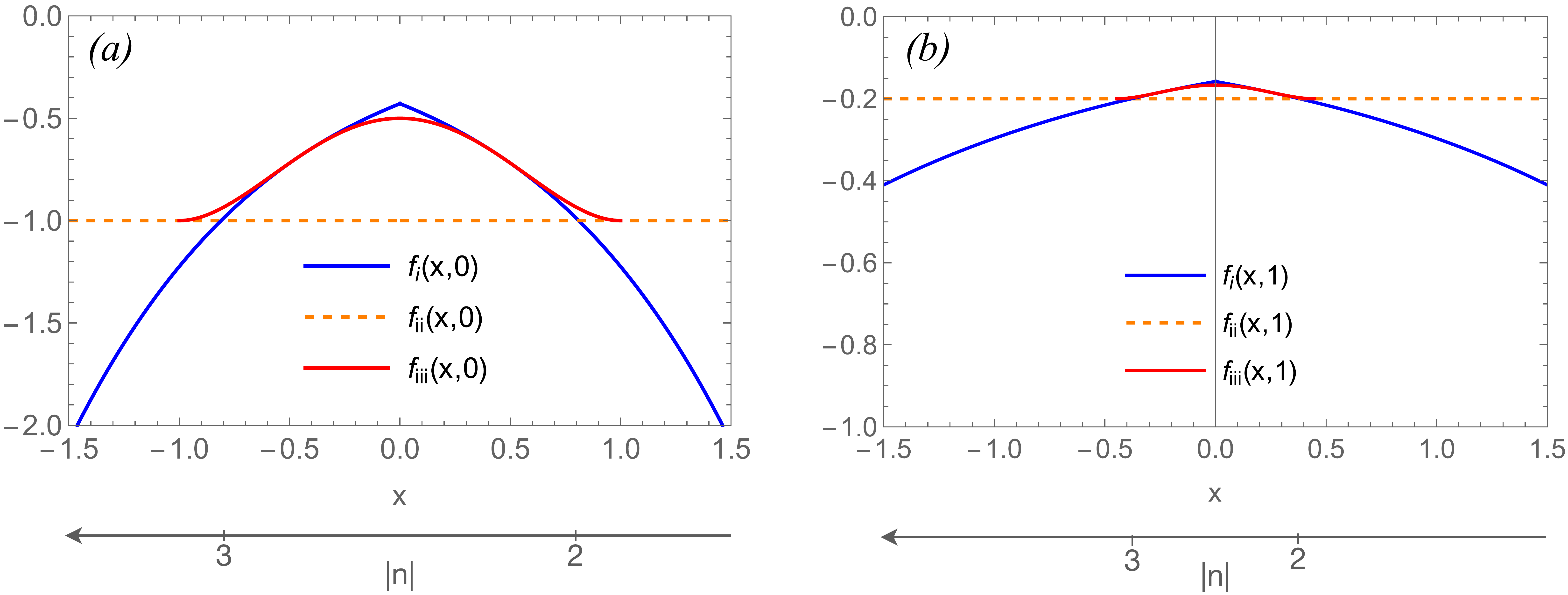}}
\centering{}
\caption{ \textbf{Landau free energy.}
 The functions $f_l (x,y)$ from Eq. (\ref{eq:f}), with $l$ =i, ii, iii, shown as functions of $x = \gamma/\sqrt{\alpha \beta}$  for
 two values of $y=\beta'/\beta$: $y=0$
 (panel (a))
  and $y=1$
   (panel (b)).
 The free energies are $F_l = (\alpha^2/\beta) f_l (x,y)$, hence the smallest $f_l (x,y)$
   determines the ground state. The states  (i) and (ii) are the ones for which the vH peaks split in 3-1 (1-3) and
   2-2 fashion, respectively. The state (iii) is an intermediate state with 2-1-1 (1-1-2) splitting.
   This intermediate state does not appear as a ground state in our model for all
    $y$, but its energy is close
    to those of (i) and (ii)  near critical $x$ of the first-order transition between the two, and it can potentially become a ground state around this $x$ if we move away from $SU(4)$-symmetric model by e.g., including
    interaction terms with inter-valley scattering.
     The relation between $x$ and $|n|$ is shown at the bottom.}
\label{prefac}
\end{figure}

  The values of $\lambda$ and the free energies for the three states,
  $F_{l} =  \frac{\alpha^2}{\beta} f_l (x,y)$,
    are
    functions of $x= \gamma/\sqrt{\alpha \beta}$ and $y = \beta'/\beta$:
    \begin{widetext}
  \bea
  && (i)  \lambda_{\mathrm{i}} = \sqrt{\frac{\alpha}{\beta}} \frac{|x|
  \pm
 \sqrt{x^2 + 7 +12y}}{7+12y},~~
    f_{\mathrm{i}} (x,y) = -
     \frac{\left(|x|
     \pm
     \sqrt{7 + x^2 +12y}\right)^2 [3(7+12y) + 2|x| (|x|
     \pm
    \sqrt{7 + x^2+12y})]}{(7+12y)^3} \nonumber \\
  &&  (ii) \lambda_{\mathrm{ii}}=\sqrt{\frac{\alpha}{\beta}} \frac{1}{\sqrt{1+4y}},~~
  f_{\mathrm{ii}} (x,y) = -
  \frac{1}{1+4y} \nonumber \\
  && (iii) \lambda_{\mathrm{iii}} = \sqrt{\frac{\alpha}{\beta}}  x,~ \lambda_{\mathrm{iii},3} = \sqrt{\frac{\alpha}{\beta}} \left(
   \sqrt{\frac{1-x^2 (1+4y)}{1+2y}} -|x|\right),~
   f_{\mathrm{iii}} (x,y) = -
   \frac{1+2x^2-x^4(1+4y)}{2(1+2y)}\,.
    \label{eq:f}
\eea
\end{widetext}
 The solution (iii) exists for $|x| \leq 1/\sqrt{1+4y}$
 and the $\pm$ sign is for positive/negative $\gamma$.
 We plot the free energy prefactors $f_{l} (x,y)$
 in  Fig. \ref{prefac}.

We see that at large $|x|$  the ground state configuration is  state (i) while for
small
$|x|$ it is state (ii).
  There is a direct first-order transition between states (i) and (ii) at some intermediate $|x|= x_\textrm{{cr}}$.
 We expect that $\alpha>0$
  between $1<|n|<4$,
  where
 the
  vH peak remains near the chemical potential,
 and
argue that $\gamma$ changes sign from positive to negative as $|n|$ increases,
   because the sign of $\gamma$ is different when the bands are empty and when they are filled.
  As a result $x$ evolves from a large positive value to a large negative one via zero upon increasing $|n|$.
   Because small $\alpha$ corresponds to large $|x|$,
  when
  the order first emerges, the system moves into state
  (i),
 and  the components of the vH peak split in
 1-3 fashion
 for positive
 $\gamma$. As $\alpha$ increases,
   $|x|$ decreases and eventually  reaches $x_{\mathrm{cr}}$, where the system undergoes a first order transition into the
   ordered state  (ii),
   for which the splitting of the components of the vH peak is 2-2.  At larger $n$,  $\gamma$ changes sign and
     its magnitude increases, while
     $\alpha$ starts decreasing. As a result, $|x|$ increases. When it reaches $x_{\mathrm{cr}}$, the system undergoes another first-order transition into the state, which gives rise to  3-1
     splitting of the components of the vH peak. Eventually the order disappears
     and all 4 components of the vH peak recombine into a single peak.
We also note that while the intermediate state (iii) is not the ground state for any $x$ and $y$, its free energy $F_{\mathrm{iii}}$
   is only slightly larger than $F_{\mathrm{i}}$ and $F_{\mathrm{ii}}$ at  $|x|$ near $x_{\textrm{cr}}$. This is particularly so   at
  large $y$ (at $x_{\textrm{cr}} \approx \sqrt{3/16y}$, $F_{\mathrm{iii}}$ is larger than $F_{\mathrm{i}} = F_{\mathrm{ii}}$ by $(\alpha^2/\beta) 1/(16y)^2$).
 Thus, it seems possible
 that the intermediate state (iii) will become the ground state once we move away from an SU(4)-symmetric model by e.g., including
    interaction terms with inter-valley scattering.  Such terms are small,
   but finite in TBG \cite{Kang2018PRX,Yuan2018,Kang2019PRL}.
   If the transition from (i) to (ii) is via the intermediate phase (iii),  there is a range of $|n|$ where the splitting of the vH peak components is 2-1-1 or 1-1-2,
   again depending on the sign of $\gamma$.
    Some indications of 2-1-1 and 1-1-2 splitting have  been found in STM for hole-doped samples~\cite{Xie2019,ali_2}.

{\bf { Cascade of transitions in BBG and RTG}} ~~~

\begin{figure}[h!]
\center{\includegraphics[width=\textwidth]{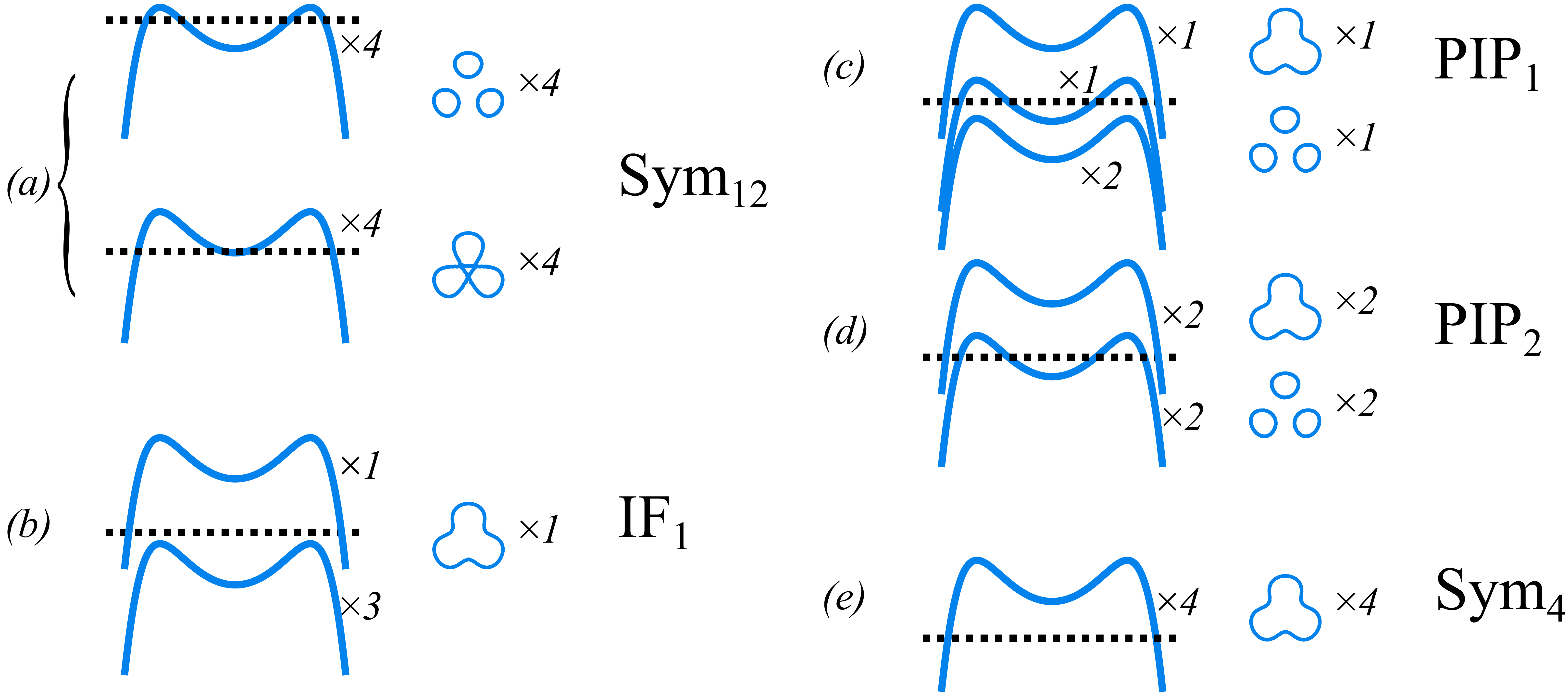}}
\begin{minipage}[h]{0.99\linewidth}
\end{minipage}
\centering{}\caption{ \textbf{The cascade of transitions in BBG/RTG.}
The notations -- 
 the same as in \cite{BBG_exp2022,Zhou2021cascade}, are
  as follows: IF$_1$ -- isospin ferromagnet, PIP$_1$ -- partially isospin-polarized phase with one large Fermi surface and one small, PIP$_2$ -- partially isospin-polarized phase with two large Fermi surface and two small, Sym$_4$ -- a symmetric phase with 4 identical large Fermi surfaces (one per isospin), Sym$_{12}$ -- a symmetric phase with 12 identical small Fermi surfaces (three per isospin).
 The states in panels
 (a)-(e)
  are
symmetric four-fold degenerate, $1-3$, $1-1-2$, $2-2$, and again symmetric four-fold degenerate,
correspondingly.
The $3-1$ and $2-1-1$ states have not been detected in~\cite{BBG_exp2022} and are not shown.   The small pockets in $1-3$ (IF$_1$) are assumed to sink below the Fermi level.  The symmetry between three small pockets in panels
 (c) and (d)
 may be broken by subleading interactions, leaving only one small pocket, as the data in \cite{BBG_exp2022}  indicate.}
\label{fig_BBG}
\end{figure}

The same analysis
 can be applied to study the cascade of phase transitions in BBG and RTG.
 In both systems, application of an electric field opens
 a gap between
 conduction and valence bands and flattens
 the fermionic dispersion near Dirac $\textbf{K}$ and $\textbf{K'}$ points~\cite{bilayer_bias}.
  Near charge neutrality, this
   creates small Fermi pockets, three near $\textbf{K}$ and three near $\textbf{K'}$. Upon doping, pockets
    merge at
    vH fillings and eventually transform into one larger pocket near $\textbf{K}$ and one near $\textbf{K'}$ (Refs. \cite{McCann2006PRL,graphene_RMP,McCann_2013,Koshino2009ABCWarping,Zhang2010ABC,Berg_21}).
 We consider the full 6-pocket model in Supplementary Discussion XI and here illustrate the behavior using a simplified
 model of fermions in two patches
 near $\textbf{K}$ and $\textbf{K'}$ with
 Hubbard intra-patch and inter-patch density-density interaction.
      In this model,
      electronic instabilities towards valley polarization, intra-valley ferromagnetism, and
      inter-valley spin and charge order all occur at the same critical
     coupling $\lambda$.  These 15 bilinears  then form an adjoint representation of SU(4)
    and are described by the
    same Landau free energy functional as in  \eqref{Lagr_main}.
     The cascade of transitions in BBG and RTG then
     matches the one in TBG
     with the only difference that some pockets may sink below the Fermi level (see Fig. \ref{fig_BBG}).
 In
a more realistic 6-patch model, the coupling
for $7$
$C_3$ symmetry preserving
order parameters with $\textbf{Q}=0$  is not the same as for order parameters  with momenta $\textbf{Q}$ close to $\textbf{K}-\textbf{K'}$.
The sequence of transition and the Fermi surface reconstruction
remain the same as
in the 2-patch model if the order develops with $\textbf{Q}=0$.

{\bf Comparison with experiments on TBG}~~~
  In   our proposed vH scenario, spin/isospin order develops at $|n| \lesssim 1$,
  when the
  four-fold degenerate vH peak approaches the Fermi level, and persists up to
   $|n| \lesssim 4$. In this range of $n$,
  the vH peak splits, but according to STM data, its components are still located near the Fermi energy, i.e., the
  enhancement of the DOS near the Fermi level persists.  At larger $|n|$, the vH peak again becomes
    four-fold
    degenerate
   and moves away from the Fermi level.
  The evolution of spin/isospin order and of its feedback on the components of the vH peak is governed
   in our theory
   by the relative strength of the prefactor of the cubic term in the Landau free energy
  (specifically, by  $x = \gamma/\sqrt{\alpha \beta}$).
  This prefactor is expressed via a convolution of three fermionic propagators and vanishes for particle-hole
  symmetry
   around
   the
   Fermi surface. In the absence of such symmetry, $\gamma$
   is non-zero.
   We conjecture that  $x$ is
   positive near $n=1$
   passes through zero at $2< n <3$, and becomes
 negative   at larger $n$ (see Supplementary Discussion V for more discussion on this).
   We then
   end up with the phase diagram  in
    Fig. \ref{cascade_pic} (f).
    There are two phase
   transitions between disordered and ordered states  at $|n| \lesssim 1$, and $|n| \lesssim 4$, and
    two transitions between different ordered phases at $|n| \lesssim 2$ and $|n| \lesssim 3$.
   Specifically, within our theory the sequence for
   the symmetry-breaking pattern and the degeneracy of the vH peak is:
   \begin{align}
n & \lesssim 1\!:  \mathrm{SU(4)} \rightarrow \mathrm{SU(3)}\!\times\!\mathrm{U(1)} :
(4,0) \rightarrow (3,1) \notag
\\
n & \lesssim 2\!:    \mathrm{SU(3)} \!\times\! \mathrm{U(1)}\rightarrow
\mathrm{SU(2)}\!\times\!\mathrm{SU(2)}\!\times\!\mathrm{U(1)}:
(3,1) \rightarrow (2,2) \notag \\
n & \lesssim\! 3:  \mathrm{SU(2)}\!\times\!\mathrm{SU(2)}\!\times\!\mathrm{U(1)}\rightarrow
\mathrm{U(1)} \!\times\! \mathrm{SU(3)}:
(2,2) \rightarrow (1,3) \notag \\
n & \lesssim 4\!:  \mathrm{U(1)}\!\times\!\mathrm{SU(3)} \rightarrow \mathrm{SU(4)}:
(1,3) \rightarrow (0,4)
\label{group_pattern}
\end{align}
where  $a$ and $b$ in  $(a,b)$ indicate the number of vH peaks
above and below $\mu$
for the case of electron doping. For hole doping
the sequence is identical, except
$a$ and $b$ in $(a,b)$ are interchanged.
If the transformations $(3,1) \to (2,2)$ and $(2,2) \to (1,3)$ occur via an intermediate phase (c), each of the two
first-order transitions around $|n|=2$ is replaced by two second-order transitions with the intermediate structure of vH peaks
$(2,1,1)$ and $(1,1,2)$.

 The  theoretical phase diagram
 agrees with the STM results \cite{Xie2019,ali_2}
   (Fig. \ref{cascade_pic} (b,d))
   including fine details, lending support to our theory.
   Note, that there is no symmetry of the phase diagram with respect to $|n|=2$, i.e. the transitions at $|n| \lesssim 1$ and $n\lesssim 3$ are different ones
    (there is an approximate symmetry with respect to $|n| = 2.5$).
   The theory also explains the
 seesaw  behavior  of electron compressibility, reported in \cite{Zondiner2020} and  shown in Fig.
 \ref{cascade_pic} (e).
Our reasoning is the following. As doping increases and the system approaches one of  transitions from the cascade,  the inverse compressibility $d\mu/dn$ decreases as
the $n$-times degenerate vH peak approaches the Fermi level, where $d\mu/dn =0$   ($n =4,3,2$, depending on the number of the transition in the cascade). After a new order develops,   one peak component crosses the Fermi level, while the other $(n-1)$ components move back from the Fermi level.  Because all vH peaks move away from the Fermi level
in a first-order transition,
$d\mu/dn$  jumps to a larger value.  As doping increases further towards the next transition from the cascade, the  $(n-1)$ – times degenerate vH peak  approaches the Fermi level, and $d\mu/dn$ again decreases  towards zero.  Then the new order develops, one peak component crosses the Fermi level, while the other $(n-2)$ components move back from it, and $d\mu/dn$ again jumps to a higher value.  This gives rise to seesaw structure of the inverse compressibility
(see Supplementary Discussion V for an example calculation of $d\mu/dn$ for one transition of the cascade).
 Because all four vH peak components remain close to the Fermi level, all four contribute to the evolution of $d\mu/dn$ between the transitions.  This is consistent with a weak dependence  of the slope of $d\mu/dn$ on the number of a transition in the cascade.

{\bf  Comparison with experiments on BBG/RTG}~~~ Measurements of inverse electronic compressibility and magnetoresistance in BBG~\cite{BBG_exp2022,Seiler2022} and RTG~\cite{Zhou2021cascade} at a finite displacement field revealed a cascade of transitions upon hole or electron doping.
 The
  fermionic structure of the two materials is almost identical,
  and for definiteness we
focus on hole-doped BBG.
 Near charge neutrality, the system is in
  the valley/spin symmetric state
  (labeled Sym$_{12}$ in Ref. \cite{BBG_exp2022} and in
   Fig. \ref{fig_BBG} (a))
   with
twelve Fermi pockets: three
spin-degenerate ones
 for each valley.
At large enough doping,
the triad of pockets for each valley and spin transforms into a
 single larger pocket,
 leaving four pockets, again valley and spin symmetric  (Sym$_{4}$ state in Ref. \cite{BBG_exp2022} and in
  Fig. \ref{fig_BBG} (e)).
The cascade of transitions happens in between these two limits, when the system develops particle-hole order that breaks valley and/or spin symmetry.
We show the sequence of transitions in the cascade in the 2-patch model in Fig. \ref{fig_BBG}.

  The authors of~\cite{BBG_exp2022,Seiler2022} detected the symmetric three intermediate phases, which they labeled IF$_1$,  PIP$_1$, and  PIP$_2$. The IF$_1$ state has one large pocket, the PIP$_2$ state has two large and
  two
  small pockets, and the intermediate PIP$_1$ state has one large and one small
   pocket.  We argue that IF$_1$ is  the state  (i) in Eq. \eqref{eq:f} with co-existing valley polarization and ferromagnetism in one valley. This order develops first and splits Fermi pockets in  1-3 fashion with one large pocket and three-fold degenerate small pockets, which may be present or sink below the Fermi level (panel
   (b)
    in Fig. \ref{fig_BBG}).  The PIP$_2$
 is the state (ii) in Eq. \eqref{eq:f} with either valley polarization or ferromagnetism in both valleys. This order
  develops at a larger magnitude of the order parameter and  splits Fermi pockets into two large and two small pockets
  (panel
  (d) in Fig.  \ref{fig_BBG}).
   In the SU(4)-symmetric case
  there are three small pockets,
  but
  their number may be
  reduced by subleading interactions.  The PIP$_1$ is the intermediate state
  (iii) in Eq. \eqref{eq:f} with one large and one small pocket
   (panel (c) in Fig.  \ref{fig_BBG}).
  Experiments
  did not reveal the 1-3 state, which is the part of our theoretical sequence. We expect this state to be present,
  but probably
   in a narrow doping range.
The spin-polarized correlated metal at the end of the cascade in Ref.~\cite{Seiler2022} is a potential candidate for the 1-3 state.
  We also note that it depends on the size of the displacement field and the splitting on which side of the van Hove energy the Fermi level ends up after the transition to the 1-3, 2-2, and 3-1 states so that more phases are possible. This provides an explanation for the additional phases observed at larger displacement field in Ref.~\cite{Seiler2022}.
  The data also show that in some range of displacement fields the system returns back to Sym$_{12}$ state in between PIP$_1$ and PIP$_2$.  In our theory, this holds if particle-hole order vanishes in this parameter range.

\section*{\uppercase{Discussion}}
In this theoretical work, we used as an input STM data for TBG, which show that upon
  electron or hole doping, one of
  the
   vH peaks in the DOS remains
   near the chemical potential in a wide range of
   fillings -- between $|n| \lesssim 1$ and $|n| \lesssim 4$.
   We analyzed a cascade of phase transitions
   imposed by evolving spin/isospin order,
   which in turn is  associated with the
   enhancement of the DOS for low-energy fermions
   due to a
    confinement of a
   vH peak
   close to
   $\mu$. We found a set of phase transitions:
    two first order transitions at  $|n| \lesssim 1$ and $|n| \lesssim 4$ between disordered and ordered states
     and two transitions at $|n| \lesssim 2$ and $|n| \lesssim 3$ between different ordered states
     with different spin/isospin order and different splitting of vH peaks.  These last  transitions can be first order
    or continuous, via  a
    narrow intermediate phase.  We argue that these transitions give rise to the
     seesaw behavior of the compressibility,
     with the jumps of $d\mu/dn$ at the first-order transitions (where we also expect hysteretic behavior of the magnetization) and continuum, but rapid changes of $d\mu/dn$ if the transition is via an intermediate phase.
     We also emphasize that in our description the minima of $d\mu/dn$ are near, but not exactly at integer $n$.

     The semi-phenomenological explanation of the cascade of transitions put forward in Ref.~\cite{Zondiner2020}
      assumes that at every transition one of 4 initially degenerate bands gets fully filled/fully emptied
       and no longer contributes to particle-hole order.
       Within  this scenario, one
 can naturally explain the emergence of insulating states at integer fillings, but one would need to explain why the four vH peaks, seemingly moving to different energies as $|n|$ increases,
   recombine into a single vH peak at $|n| \lesssim 4$, as STM
 data show,
  and would also need to explain why the measured slope of $d\mu/dn$ does not scale inversely with the number of remaining peak components.
    We discuss this scenario in some detail in the Supplementary Discussion VIII. Interestingly, it yields
  the same
   ordered states
      as in our SU(4) scenario.

  There is an element of phenomenology in our approach as well.
     Namely, we departed from a metal, associated the emergence
     of spin/isospin order with a vH singularity, and associated the cascade of transitions with near-integer $|n|$
     based on STM data rather than on microscopic calculations.  The confinement of transitions to integer $|n|$ and the emergence of insulating phases around these $|n|$ are
      most likely
     strong-coupling phenomena.
 We note in this regard that the SU(4)-symmetric  Landau free energy, on which our results and the results of
      Ref. \cite{Zondiner2020}  are based upon, is
       in fact
       generic, and while we derived it from the specific microscopic itinerant model of interacting fermions with $\mu$ near the vH singularity, the same expression can be obtained in a strong coupling limit, where
       the bands are assumed to be nearly completely flat
        \cite{Kang2020RG,Zondiner2020,Khalaf2020soft,BernevigTBG4,Kang2021PRL,LEDWITH2021168646}, and their internal structure does not play a role. Within
        the strong-coupling scenario,
         the STM peaks, which we interpreted as van Hove peaks, are treated as the peaks corresponding to flat bands.
         In either scenario, the
         Luttinger theorem states that
         the splitting due to spin/isospin orders
         can
         lead
         to the formation
         of insulating states only at integer fillings.
          A similar conclusion that a symmetry-breaking  occurs at a non-integer filling due to vH physics and
          gives rise to an insulating behavior near integer $n$ has been reached in Ref.~\cite{PhysRevLett.127.196401}.
In a recent
 experimental study~\cite{Nadj_Perge_cascade}
the authors argued  that the cascade of transitions in TBG is present in a range of twist angles, even when there are no insulating states near integer fillings. These results lend further support to our van Hove-based scenario of the cascade  of phase transitions in TBG.

Our theory also describes the cascade of phase transitions, detected in compressibility and magnetoresistance measurements
in BBG and RTG under a displacement field. These systems have small Fermi pockets near $\textbf{K}$ and $\textbf{K'}$, which undergo a set of transitions around the vH doping. We argue that the splitting of the pockets in different phases in the cascade is the same as in TBG and is caused by the same set of valley and spin orders.   The similarity of the cascade phases in TBG and BBG/RTG is quite striking given that BBG/RTG are substantially less correlated than TBG because an application of the displacement field flattens the dispersion near $\textbf{K}$ and $\textbf{K'}$, but the full bandwidth remains the same as in the original non-twisted bilayer graphene.  We believe that the similarity is
an indication that the structure of particle-hole order in all three systems and
 the structure of
the accompanied splitting of the electron bands can be understood already by analyzing what are the leading  instabilities of a doped metal with valley and spin degrees of freedom.
 A strong coupling approach is certainly needed for the description of how  in TBG the order creates an insulating behavior near integer fillings.

 \paragraph*{\bf{Acknowledgment}}~~~We thank E. Berg, A. Cherman, Z. Dong, R. Fernandes, F. Guinea, J. Hoffman, S. Ilani, P.
  Jarillo-Herrero,
 E. K\"onig,
 C. Lewandowski,
  L. Levitov, Y. Oreg,
 H. Polshyn,
 G. Tarnopolsky, O. Vafek, A. Vishwanath, A. Yazdani, A. Young, and E. Zeldov for fruitful discussions. We are indebted to
   S. Ilani, K. Nuckolls, A. Yazdani, and  U. Zondiner  for sharing their data with us. The work by D.V.C and A.V.C. was supported by
 U.S. Department of Energy, Office of Science, Basic Energy Sciences, under Award No. DE-SC0014402.
 Y.W. was supported by NSF under award number DMR-2045781.
 D.V.C. gratefully acknowledges support from  Doctoral Dissertation  and Larkin Fellowships at the University of
 Minnesota.

\paragraph*{\bf{Data availability}}
Data will be kept in UMN database and will be available upon request.

\paragraph*{\bf{Author contributions}}
D.V.C. performed analytic calculations with an input from L.C.,  Y.W., and A.V.C.  D.V.C. and A.V.C. wrote the first draft. All authors discussed the results and their relation to experiments, and contributed to writing the manuscript.

\paragraph*{\bf{Competing interests}} The authors declare no competing interests.

\newpage

\bibliography{biblio_cascade}
\bibliographystyle{naturemag}

\clearpage

\begin{center}
\textbf{\large Supplemental Material}
\end{center}
\setcounter{section}{0}
\setcounter{equation}{0}
\setcounter{figure}{0}
\setcounter{table}{0}
\setcounter{page}{1}
\makeatletter
\renewcommand{\thesection}{S\arabic{section}}
\renewcommand{\theequation}{S\arabic{equation}}
\renewcommand{\thefigure}{S\arabic{figure}}

\section*{Supplementary Discussion I: The ground state of an SU(4)-symmetric free energy}
\label{sec:minimaSUN}

In the main text we discuss an effective  model of fermions near six (or twelve) van Hove points in twisted  bilayer graphene (TBG). The model consists of 4 sets of interacting fermions (2 spin and 2 valley isospin variables).
We argued previously~\cite{Chichinadze2021su4} that in both cases (six or twelve van Hove points) there are 15 particle-hole order parameters (bilinear combinations of fermions)
with nearly equal
 attractive couplings.
  The couplings for other order parameters are either repulsive or smaller by magnitude.
  More precisely, the set of 15 consists of
  two subsets of 7 and 8 bilinears. The couplings within each subset are identical ($\lambda_7$ for the first subset, $\lambda_8$ for the second).  The couplings $\lambda_7$ and $\lambda_8$ are not identical, but are close to each other, and in our analysis we treated them as equal.
       These 15 fermionic bilinears
      then
      form an adjoint representation of the SU(4) group.  Here we analyze in some detail the Landau free energy of the corresponding SU(4) model and obtain the structure of the ordered state for different parameters.  For completeness, we also analyze the structure of the order in SU(N) models with $N=3$,
      which in our case are effective models of fermions with 3  degenerate bands  and $N^2-1 =8$
       particle-hole bilinears with equal couplings (see Supplementary Discussion VI below).   It is convenient to consider
       the general SU(N) case for the free energy and specify $N=4$ or $N=3$ in the next section.

The free energy of a system of
$N^2-1$  degenerate particle-hole order parameters $\phi_i$ is, to
fourth order   in $\phi_i$
\begin{equation}
\begin{gathered}
\mathcal{F} = - \frac{\alpha}{2} \mathrm{Tr}[\hat{\Phi}^2] + \frac{\gamma}{3} \mathrm{Tr}[\hat{\Phi}^3] + \frac{\beta}{4} \mathrm{Tr}[\hat{\Phi}^4] +\frac{\beta'}{4} \mathrm{Tr}[\hat{\Phi}^2]^2.
\end{gathered}
\label{Lagr}
\end{equation}
Here $\hat{\Phi} = \sum_j^{N^2-1} \phi_j T^j$, where $T^j$ are the generators of the group SU(N). Note, that the prefactors in the free energy are defined in a slightly different way than in \cite{Chichinadze2021su4}.
 The
   order-parameter fields $\phi_j$ are expressed via fermionic bilinears as
\begin{equation}
\phi_j
\sim
f^{\dagger} T^j f
\end{equation}
where $f^{
\dagger)
}$  and $f$ are electronic creation and annihilation operators.
The momentum-transfer between electrons can be zero or finite depending on which order $\phi_j$ corresponds to.

This
effective model can be straightforwardly obtained by departing from the Hamiltonian for $N$ species with equal dispersion and 4-fermion interactions, selecting $N^2-1$ particle-hole bilinears,
 applying a Hubbard-Stratonovich transformation,
and integrating out the fermions.
Note that in the effective model we considered previously~\cite{Chichinadze2021su4}, the dispersion of all fermion species was non-degenerate for different valleys so that the SU(4) symmetry was only approximate.
 The $\beta'$-term in (\ref{Lagr}) does not appear in the expansion of the logarithm in the Hubbard-Stratonovich formalism, but
is  allowed on general grounds and in practice is generated in the renormalization group flow \cite{Shrock2010symmetry}.
This term and the $\beta$ term are
 the two independent quartic terms for $N= 4$. For $N=3$, $\mathrm{Tr}[\hat{\Phi}^2]^2 =2\mathrm{Tr}[\hat{\Phi}^4] $, in which case
 the two terms are equivalent and we can  $\beta'$ into $\beta$.

In the Hubbard-Stratonovich formalism the sign of $\gamma$ is determined by the sign of a
one-loop diagram
with
three fermion-boson vertices and
 three propagators of low-energy fermions. The diagram vanishes in the limit when there is  a particle-hole symmetry
 around the Fermi surface,
 but is non-zero in a generic case. For definiteness, below we set $\gamma >0$.
 The extension to the case $\gamma <0$ changes the overall sign of the order parameter.

 Because $\hat{\Phi}$ is in the adjoint representation of the SU(N) group, it can be represented by a traceless matrix~\cite{Hamermesh}. It is convenient to apply a unitary transformation and analyze $\hat{\Phi}$ in the diagonal basis, where it takes the form
\begin{equation}
\hat{\Phi} = \mathrm{diag}(\lambda_1, \lambda_2, ..., \lambda_{N-1}, -(\lambda_1 + ... + \lambda_{N-1})).
\end{equation}
In this  representation the free energy (\ref{Lagr}) reads
\begin{align}
\mathcal{F} &= - \frac{\alpha}{2} \left( \sum_j^{N-1} \lambda_j^2 + (\sum_j^{N-1} \lambda_j)^2 \right) + \frac{\gamma}{3} \left( \sum_j^{N-1} \lambda_j^3 - (\sum_j^{N-1} \lambda_j)^3 \right)  \notag\\
&+ \frac{\beta}{4} \left( \sum_j^{N-1} \lambda_j^4 + (\sum_j^{N-1} \lambda_j)^4 \right) + \frac{\beta'}{4}\left( \sum_j^{N-1} \lambda_j^2 + (\sum_j^{N-1} \lambda_j)^2 \right)^2.
\label{Lagr2}
\end{align}
Note that the first three terms in (\ref{Lagr2}) have the form of
\begin{equation}
\frac{\varkappa_n}{n} \left( \sum_j^{N-1} \lambda_j^n + (-1)^{n} (\sum_j^{N-1} \lambda_j)^n  \right),
\end{equation}
where $\varkappa_n\in \{\alpha,\beta,\gamma\}$. Below we explicitly minimize the free energy (\ref{Lagr2}) for  $N=4,3,2$.
 For definiteness we assume that the prefactor  $\alpha$ in (\ref{Lagr2}) is positive and analyze energy minimization
  for finite $\lambda_j$.  The combination of
  stationary non-zero $\lambda_j$  determines the type of order
  in terms of fermion bilinears $\phi_j \sim \langle f^{\dagger} T^j f \rangle$ and also determines the shift of the van Hove peaks of reconstructed energy bands.


For $N=4$  the order parameter matrix $\hat \Phi$ in the diagonal basis is
\begin{equation}
\hat{\Phi} = \mathrm{diag}(\lambda_1, \lambda_2, \lambda_{3}, -(\lambda_1 + \lambda_2 + \lambda_{3})).
\end{equation}
The free energy (\ref{Lagr2}) is then a function of three parameters $\lambda_{1,2,3}$.
The values of these parameters are determined by the condition that the free energy is at a minimum.
We
analyze the
solutions of $\frac{\partial F}{\partial \lambda_l} = 0$. In explicit form
\begin{widetext}
\begin{align}
\frac{\partial F}{\partial \lambda_l} & =  - \alpha \left(  \lambda_l + \sum_{j} \lambda_j \right) + \gamma \left[ \lambda_l^2 - \left(\sum_{j} \lambda_j\right )^2  \right] + \beta \left[ \lambda_l^3 +\left(\sum_{j} \lambda_j\right)^3  \right] +\beta' \left[ \sum_{j} \lambda_j^2 + \left(\sum_{j} \lambda_j\right )^2  \right]\left( \lambda_l + \sum_{k } \lambda_k \right) = 0
\label{extr}
\end{align}
In our previous work \cite{Chichinadze2021su4}  we set  $\gamma = \beta' =0$ and found two solutions:
$\lambda_1 = \lambda_2 = - \lambda_3 = (\alpha/\beta)^{1/2}$ and $\lambda_1 = \lambda_2 = \lambda_3 = (\alpha/(7\beta))^{1/2}$.  For the first solution a 4-fold
 degenerate van Hove level splits into two doubly degenerate levels,  for the second one the van Hove level splits into
   one 3-fold degenerate level and one single level.
    The first solution has lower free energy, hence the only option at $\gamma = \beta'= 0$ is 2-2 splitting.

For a generic case when $\gamma$ and $\beta'$ are non-zero,  we
 subtract Eq. (\ref{extr}) for $l=2$ from that for  $l=1$
 and obtain
\begin{align}
0=(\lambda_1 - \lambda_2)&\left[ - \alpha + \gamma (\lambda_1 + \lambda_2) + \beta (\lambda_1^2 + \lambda_1 \lambda_2 + \lambda_2^2) +\beta'(\sum_j\lambda_j^2+(\sum_j\lambda_j)^2) \right].
 \label{nn}
\end{align}
We see that either $\lambda_1 = \lambda_2$ or the
$\lambda$'s have to satisfy the quadratic equation in square brackets in (\ref{nn}).
 Below  we search for the solutions of (\ref{extr})
with the constraint $\lambda_1 = \lambda_2$.  It is straightforward to verify that enforcing instead the  condition set by the quadratic equation in (\ref{nn}) ultimately leads to the same solutions with  permuted $\lambda_j$.

Let us label $\lambda_1 = \lambda_2 = \lambda$. The remaining two equations in (\ref{extr}) are
\begin{align}
0&=2(\lambda + \lambda_3) \left( -\alpha - 2\gamma \lambda  + \beta (4\lambda^2 + 2\lambda \lambda_3 + \lambda_3^2) +2\beta'(3\lambda^2+\lambda_3^2+2\lambda\lambda_3)\right)  \notag \\
0&=(\lambda - \lambda_3) \left( -\alpha + \gamma (\lambda + \lambda_3) + \beta (\lambda^2 + \lambda \lambda_3 + \lambda_3^2) +2\beta'(3\lambda^2+\lambda_3^2+2\lambda\lambda_3)\right)
\end{align}
The upper equation follows from (\ref{extr}) for $l=3$ and the lower one is obtained by subtracting
the equation for $l=3$ from
 that
 for $l=1$.
There are three possible solutions of these equations:
(i) $\lambda = \lambda_3$, where $\lambda$ is determined by the second bracket in the first equation;
 (ii) $\lambda = -\lambda_3$, where $\lambda$ is determined by the second bracket in the second equation; (iii)
   $\lambda$ and $\lambda_3$ are different and are  determined by second brackets in both equations.
Let us consider those conditions one by one.
\end{widetext}

(i)  $\lambda = \lambda_3$.
Solving for $\lambda$, we obtain
\begin{equation}
\lambda = \frac{\gamma + \sqrt{\gamma^2 +  \alpha (7\beta+12\beta')}}{7\beta+12\beta'}
\label{eq:lambdai}
\end{equation}
The free energy at the minimum is given by
\begin{equation}
 \mathcal{F} =
  \frac{\alpha^2}{\beta} f_i(x,y),
\end{equation}
where $x = \frac{\gamma}{\sqrt{\alpha \beta}}$, $y=\beta'/\beta$, and
\begin{align}
&f_i(x,y) = \label{eq:fi} \\ \notag
& -
\frac{(x +\! \sqrt{7 + x^2 +12y})^2 \left[3(7+12y) + 2x \left(x + \!\sqrt{7 + x^2+12y} \right) \right]}{(7+12y)^3}.
\end{align}
If $\gamma<0$, the solution with a minus sign in front of the square roots in Eqs.~\eqref{eq:lambdai}+\eqref{eq:fi} corresponds to a minimum.

(ii) $\lambda = -\lambda_3$. In this case
\beq
\lambda = \sqrt{\frac{\alpha}{\beta(1+4y)}},
\eeq
and
\begin{equation}
\mathcal{F} =
  \frac{\alpha^2}{\beta} f_{ii} (x,y).
\end{equation}
where
\begin{equation}
f_{ii}(x,y)=
- \frac{1}{1+4y}
\end{equation}
Note that $f_{ii}$ is independent of $x$.

(iii) $\lambda$ and $\lambda_3$ are determined by the two
quadratic equations:
\begin{align}
0&= -\alpha + \gamma (\lambda + \lambda_3) + \beta (\lambda^2 + \lambda \lambda_3 + \lambda_3^2) +2\beta'(3\lambda^2+\lambda_3^2+2\lambda\lambda_3) \notag\\
0&= -\alpha - 2\gamma \lambda  + \beta (4\lambda^2 + 2\lambda \lambda_3 + \lambda_3^2) +2\beta'(3\lambda^2+\lambda_3^2+2\lambda\lambda_3)
\end{align}
Subtracting one from the other we obtain
\begin{equation}
(3 \lambda + \lambda_3)(\gamma - \beta \lambda) = 0.
\end{equation}
The solution $\lambda_3 = -3 \lambda$ brings us back to case (i).
 The other solution is
  \bea
  \lambda &=& \frac{\gamma}{\beta} \nonumber \\
\lambda_3 &=& - \frac{\gamma}{\beta} \pm \sqrt{\frac{\frac{\alpha}{\beta} - \frac{\gamma^2}{\beta^2}\left(1+4\frac{\beta'}{\beta}\right)}{1+2\frac{\beta'}{\beta}}}
\label{lambdaiii}
\eea
This solution exists only for $|x| \le 1/\sqrt{1+4y}$.
The corresponding free energy is
\begin{equation}
\mathcal{F} =
 \frac{\alpha^2}{\beta} f_{iii} (x,y)
\end{equation}
where
\begin{equation}
f_{iii} (x,y)=
- \frac{1+2x^2-x^4(1+4y)}{2(1+2y)}
\end{equation}
Note that the free energy
is the same
for both signs in front of the square root  in Eq. (\ref{lambdaiii}).

The  solution (iii) merges with the solution (ii) at
   $|x| = 1/\sqrt{1+4y}$, where   $\lambda_3 = -\lambda$ and  with the solution  (i) at a smaller  $|x| = 1/\sqrt{5+12y}$ where $\lambda_3 = \lambda$ or $-3\lambda$, depending on the sign in front of the square root in (\ref{lambdaiii}).

We plot  $f_l (x,y)$ ($l=i,ii,iii$) in Fig. 2 of the main text as function of $x$ for $y=0$ and $y=1$.
 In both cases, for large $x$ (small positive $\alpha$ at a non-zero $\gamma$), the
  smallest free energy is for the state (i), and for small $x$ (small $\gamma$ and a finite $\alpha$) the
   smallest free energy is for  the state (ii).
   The intermediate state (iii) is never a minimum of $\mathcal{F}$.
     This means that there is a direct first-order transition between states (i) and (ii) at some critical $x_{cr}$.
      This result holds for all $y$.
      Still, $f_{iii} (x,y)$ is
      close to $f_{i} (x,y)$ and $f_{ii} (x,y)$ near where they become equal,
      particularly
      at large $y$.  It is then entirely possible that  $f_{iii}$ becomes
      the ground state near $x_{cr}$ if we move the system  away from SU(4) symmetry by, e.g., adding interaction terms which scatter between valleys.
       In this case the transition from state (i) at large $x$ to state (ii) at small $x$ becomes a continuous one via an intermediate phase.

\section*{Supplementary Discussion II: Symmetry properties}

 (i)  $\lambda = \lambda_3$.
 In  the diagonal basis, the matrix $\hat\Phi$ is
\begin{align}
\hat\Phi = \mathrm{diag} (\lambda, \lambda, \lambda, -3\lambda),
\end{align}
The remnant symmetry of this state is $\mathrm{SU}(3)\times \mathrm{U}(1)$, which is
a
 subgroup of the original SU(4) symmetry group.
Here,  SU(3) corresponds to symmetry transformations in the subspace of the first three components of $\hat\Phi$ and
  U(1) to a
  relative
  phase variation
  between
  the  last
  and the first three
  components.
  Up to rotations of the overall phase,
   the  SU(3) and U(1) generators are given by  block-diagonal matrices
\begin{align}
U_{\mathrm{SU(3)}} = \mathrm{diag} (e^{i \alpha_i T^i}, 1), ~~~ U_{\mathrm{U(1)}} =\mathrm{diag} (\mathbb{1}_3,
e^{i\theta})
\end{align}
where $T^{i}$, $i=1
,\ldots, 8$ are eight Gell-Mann matrices, and $\mathbb{1}$ is a $3\times 3$ identity matrix.
It can be explicitly checked that transformation matrices $U_{\mathrm{SU(3)}}$ and $U_{\mathrm{U(1)}}$ commute with the order parameter matrix $\hat{\Phi}$, therefore, the symmetry of the ground state is SU(3)$\times$U(1).  \\

 (ii) $\lambda = -\lambda_3$.
The matrix $\hat\Phi$ is
\begin{align}
\hat\Phi=\mathrm{diag}(\lambda,\lambda,-\lambda,-\lambda)
\end{align}
The remnant symmetry of this state is
 SU(2)$\times$SU(2)$\times$U(1),
 where the two SU(2)'s correspond to rotations within the subsets of the first two and the last two components of $\hat\Phi$,
  and U(1) corresponds to a rotation of one subset relative to the other.
  They can be represented by
  \begin{align}
U_{\mathrm{SU(2)}} &= \mathrm{diag} (e^{i \alpha_i \sigma^i}, \mathbb{1}_2)\notag\\
U'_{\mathrm{SU(2)}} &= \mathrm{diag} (\mathbb{1}_2,e^{i \alpha_i \sigma^i}) \notag\\
U_{\mathrm{U(1)}} &=\mathrm{diag} (\mathbb{1}_2, e^{i\theta}\mathbb{1}_2)\,,
\end{align}
where $\sigma^i$ are the Pauli matrices.

(iii) $\lambda \neq   \pm \lambda_3$.
The matrix $\hat\Phi$ is
\begin{align}
\mathrm{diag} (\lambda, \lambda, -\lambda_3, -2\lambda+ \lambda_3)
\end{align}
The remnant symmetry of this state is
 SU(2)$\times$U(1)$\times$U(1),
 where SU(2) correspond to rotations within the subsets of the first two  components and the two U(1)'s
  corresponds to independent,
  relative
   rotations of the third and the fourth components,
   e.g.,
  \begin{align}
U_{\mathrm{SU(2)}} &= \mathrm{diag} (e^{i \alpha_i \sigma^i}, \mathbb{1}_2)\notag\\
U_{\mathrm{U(1)}} &= \mathrm{diag} (\mathbb{1}_2,e^{i \theta},1) \notag\\
U'_{\mathrm{U(1)}} &=\mathrm{diag} (\mathbb{1}_2, 1, e^{i\theta})\,.
\end{align}

\section*{Supplementary Discussion III: Relaxing SU(4) symmetry}
As we mentioned at the beginning, the 15 order parameters with near-equal, attractive couplings near van Hove filling consist of two subsets with seven intra-valley, $Q=0$ order parameters and eight inter-valley, $Q\neq 0$ order parameters.
   The set with seven order parameters is described by the order parameter matrix $\hat\Phi_7=\sum \varphi^{(7)}_{ij} \sigma_i\tau_j$ with $i=0,\ldots,3$, $j=0,3$, excluding $i=j=0$, and $\sigma_i$ ($\tau_j$) are Pauli matrices for (iso-)spin degrees of freedom, where $\sigma_0=\tau_0=\mathbb{1}_2$. The set with eight order parameters is described by $\hat\Phi_8=\sum \varphi^{(8)}_{ij} \sigma_i\tau_j$ with $i=0,\ldots,3$ and $j=1,2$~\cite{Chichinadze2021su4}.
 There is a single coupling constant within each set ($\lambda_7$ and $\lambda_8$, respectively).  In the microscopic model that we used to derive the free energy, $\lambda_7$ and $\lambda_8$ are close in magnitude.
  In our analysis above, we neglected the difference between these two couplings, in which case the  free energy is SU(4)-symmetric.

  Here we briefly analyze what happens if we do not treat $\lambda_7$ and $\lambda_8$ as equal and instead assume that
   one of the two is larger and the order is formed within either the set of 7 or the set of 8.
    Keeping only one of the sets, integrating out fermion fields, and approximating the bare Green's functions for all spins and isospins to be equal, we obtain again the free energy in the form
\begin{align}
    \mathcal F_\mu=- \frac{\alpha}{2} \mathrm{Tr}[\hat{\Phi}_\mu^2] + \frac{\gamma}{3} \mathrm{Tr}[\hat{\Phi}_\mu^3] +
\frac{\beta}{4} \mathrm{Tr}[\hat{\Phi}_\mu^4]
\label{eq:F78}
\end{align}
with $\mu=7,8$. Here we neglect any symmetry-allowed terms that are not produced by integrating out the fermions
 and appear as
 higher-order effects.

Diagonalization of $\hat\Phi_\mu$ shows that \begin{align}
    \hat\Phi_7 &\sim\mathrm{diag}[\lambda_1,\lambda_2,\lambda_3,-(\lambda_1+\lambda_2+\lambda_3)] \\
    \hat\Phi_8 &\sim\mathrm{diag}[\lambda_1,-\lambda_1,\lambda_2,-\lambda_2]\,.
    \label{phi_matr_7_8}
\end{align}
Using this in Eq.~\eqref{eq:F78}, we obtain
 for $\mathcal F_7$
the same expression
as in the SU(4) symmetric case. Thus,
  within the set of 7 intra-valley $Q=0$ order parameters one obtains the same ordered states as in SU(4)-symmetric model,
   and hence the same sequence of splittings   of vH peak components (1-3 (3-1) and 2-2) and
the same cascade of transitions.
 The remaining symmetries of the ordered states differ from the ones in the case with full SU(4) symmetry since SU(2)$\times$SU(2)$\times$U(1) is already a subgroup of SU(4). In the reduced symmetry case the first transition ($n=1$) gives SU(2)$\times$U(1) residual symmetry, the second transition ($n=2$) leads to SU(2) residual symmetry, which can be either spin or valley, depending on the realization of the ground state.

 For the set of 8 inter-valley order parameters $\mathrm{Tr}[\hat{\Phi}_\mu^3]$ vanishes.  Then the ordered state gives rise to only $2-2$ splitting of vH peaks. There is no phases with 1-3 (3-1) splitting and, hence, no cascade of transitions.

\section*{Supplementary Discussion IV: The ground state of an SU(3)-symmetric free energy}
\label{sec:SU3min}

We next consider the case of an SU(3)-symmetric free energy.
Such a
 symmetry of fermionic bilinears
can
emerge
when there are three degenerate bands near van Hove points (this is the case when
  the dispersion of one of the initially 4 degenerate bands shifts such that the whole band moves away from
   the chemical potential).  Like we said, for the $N=3$ case one can set $\beta'=0$ without losing generality.

   In the SU(3)-symmetric case the order parameter matrix reads
\begin{equation}
\hat{\Phi}
 = \mathrm{diag}(\lambda_1, \lambda_2,  -(\lambda_1 + \lambda_2))
\end{equation}
and the conditions for extrema are
\begin{align}
-\alpha (2 \lambda_1 + \lambda_2) + \gamma ( \lambda_1^2 - (\lambda_1 + \lambda_2)^2 ) + \beta ( \lambda_1^3 + (\lambda_1 + \lambda_2)^3 ) &= 0, \\
-\alpha (2 \lambda_2 + \lambda_1) + \gamma ( \lambda_2^2 - (\lambda_1 + \lambda_2)^2 ) + \beta ( \lambda_2^3 + (\lambda_1 + \lambda_2)^3 ) &= 0.
\label{extrSU3}
\end{align}
Subtracting  one equation from the other,  we obtain that either $\lambda_1 = \lambda_2$, or $-\alpha - \gamma (\lambda_1 + \lambda_2) + \beta (\lambda_1^2 + \lambda_2^2 + \lambda_1 \lambda_2) =0$.
 In the first case, $\hat{\Phi}
  = \mathrm{diag}(\lambda, \lambda,  -2\lambda)$, where
 \beq
 \lambda = \frac{\gamma \pm \sqrt{12 \alpha \beta + \gamma^2}}{6\beta}.
 \label{lambdaSU3}
 \eeq
  The free energy is
  \beq
 F = -\frac{\alpha^2}{\beta} g(x),
  \eeq
 with
 \beq
 g(x) = \frac{\left( x \pm \sqrt{12+x^2} \right)^2 \left(18+x \left( x \pm \sqrt{12+x^2} \right) \right)}{432}.
 \label{SU3free}
  \eeq
 The choice of $\pm$  in \eqref{lambdaSU3},\eqref{SU3free}  depends on the sign of $\gamma$.

 We analyzed the second condition,  $-\alpha - \gamma (\lambda_1 + \lambda_2) + \beta (\lambda_1^2 + \lambda_2^2 + \lambda_1 \lambda_2) =0$,  and found that it yields $\lambda_2 =-2 \lambda_1$, i.e., $\hat{\Phi}
  = \mathrm{diag}(\lambda, -2\lambda,  \lambda)$. This is the same $\hat{\Phi}'$ as above,
   up to permutations of the components.
  Accordingly, $\lambda$ and the free energy are the same as in \eqref{SU3free}.

  We argue therefore that in the SU(3) model the ordered state is the same for all $|x|$.  This is the
   key distinction from the SU(4) model, where the ordered state changes between large and small $|x|$.

  The remnant symmetry of the system
    in the ordered state
    is $\mathrm{SU}(2)\times \mathrm{U(1)}$. Here,  SU(2) corresponds to symmetry transformations for the first two components of $\hat{\Phi}
  = \mathrm{diag}(\lambda, \lambda, -2\lambda)$ and
 U(1) corresponds to
 a relative phase variation
 between the first two and
 the last  components.

\section*{Supplementary Discussion V: Example calculation for a van Hove peak}
\begin{figure}[h]
\centering{}
\includegraphics[width=0.48\linewidth]{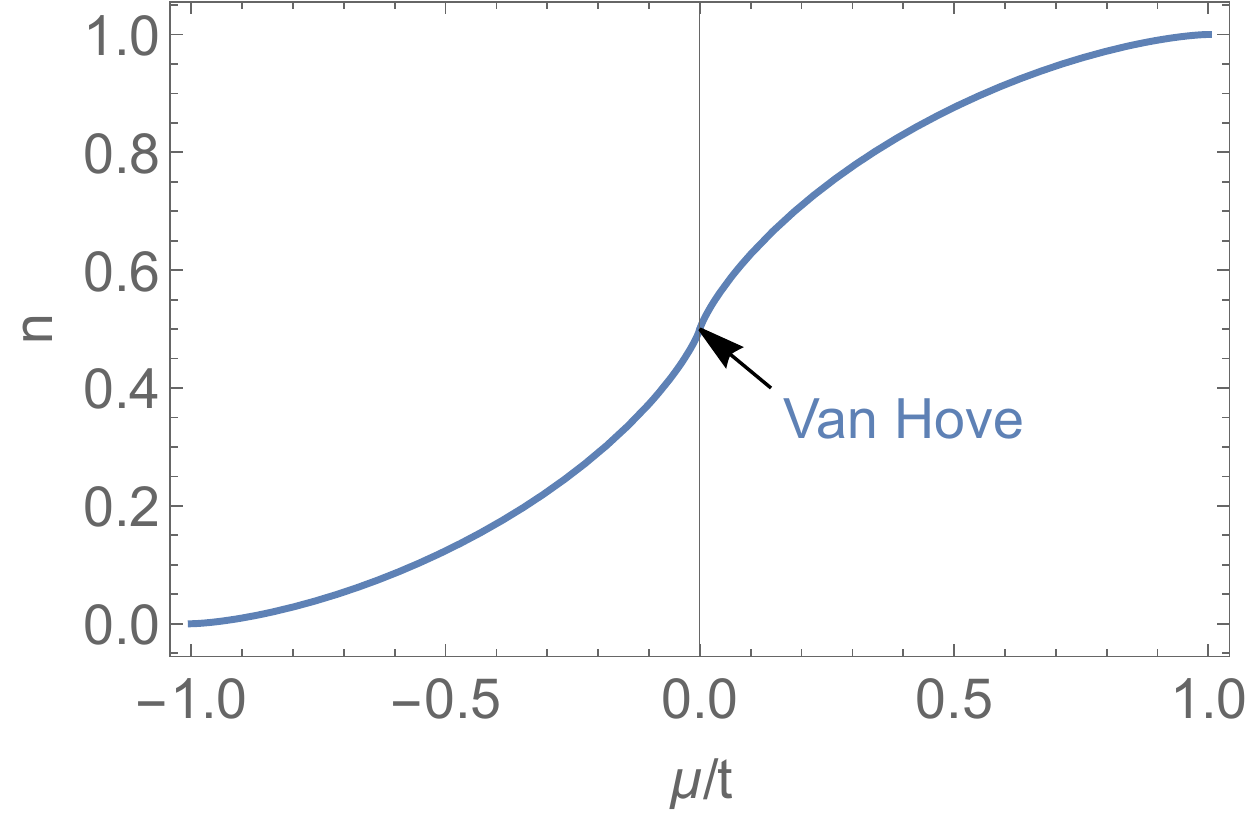}
\includegraphics[width=0.5\linewidth]{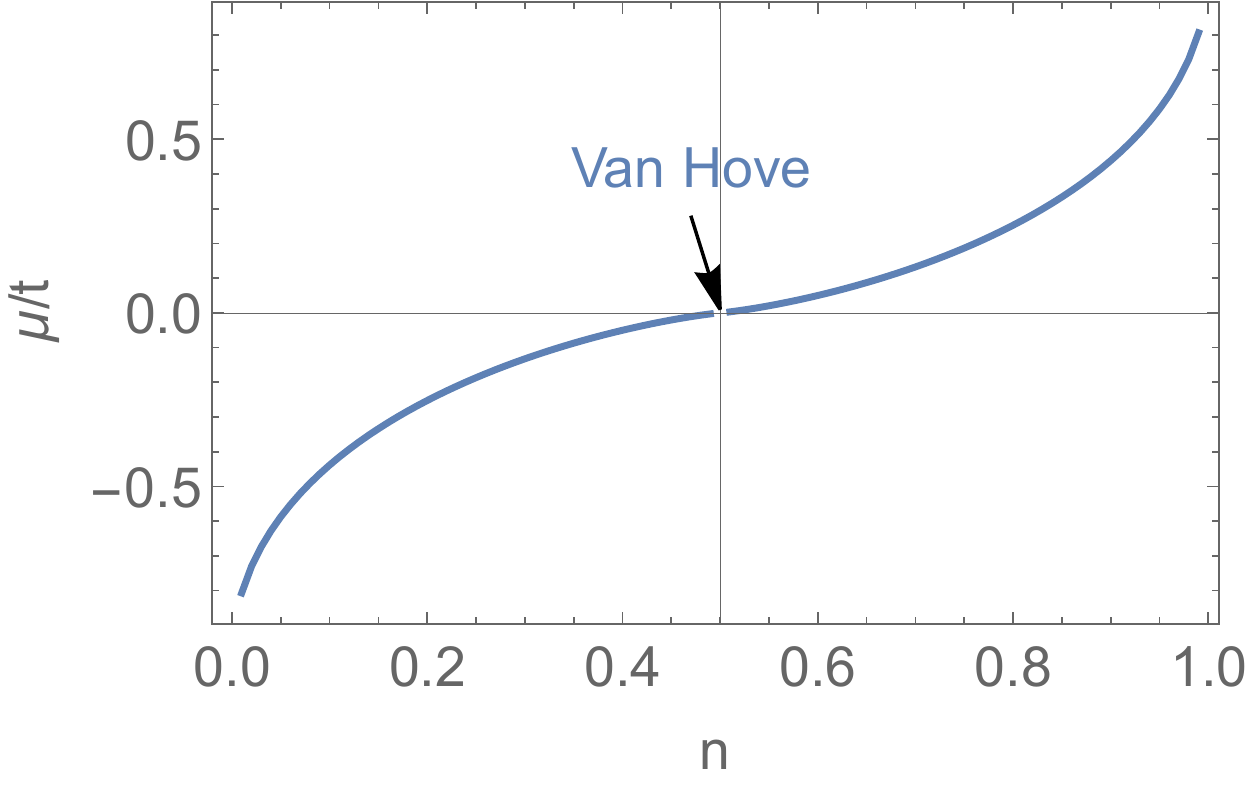}\\
\includegraphics[width=0.49\linewidth]{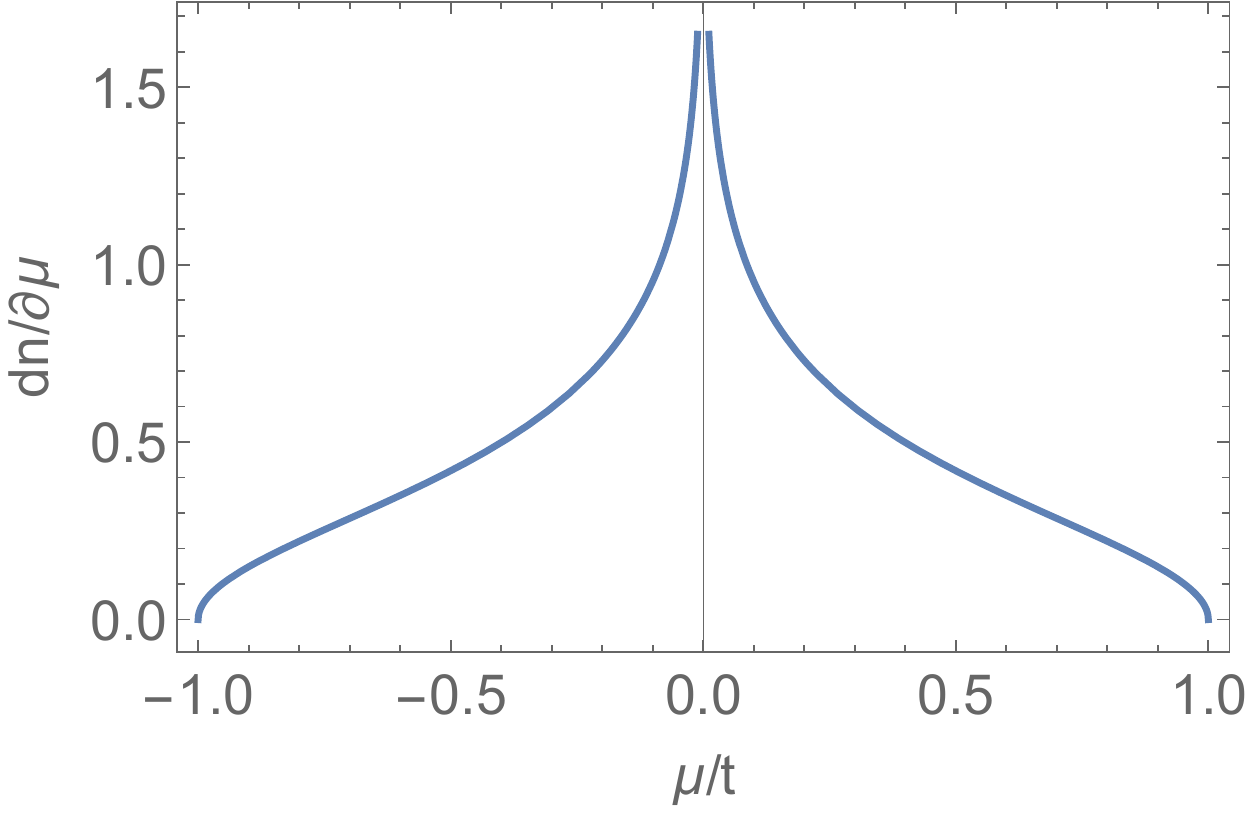}
\includegraphics[width=0.49\linewidth]{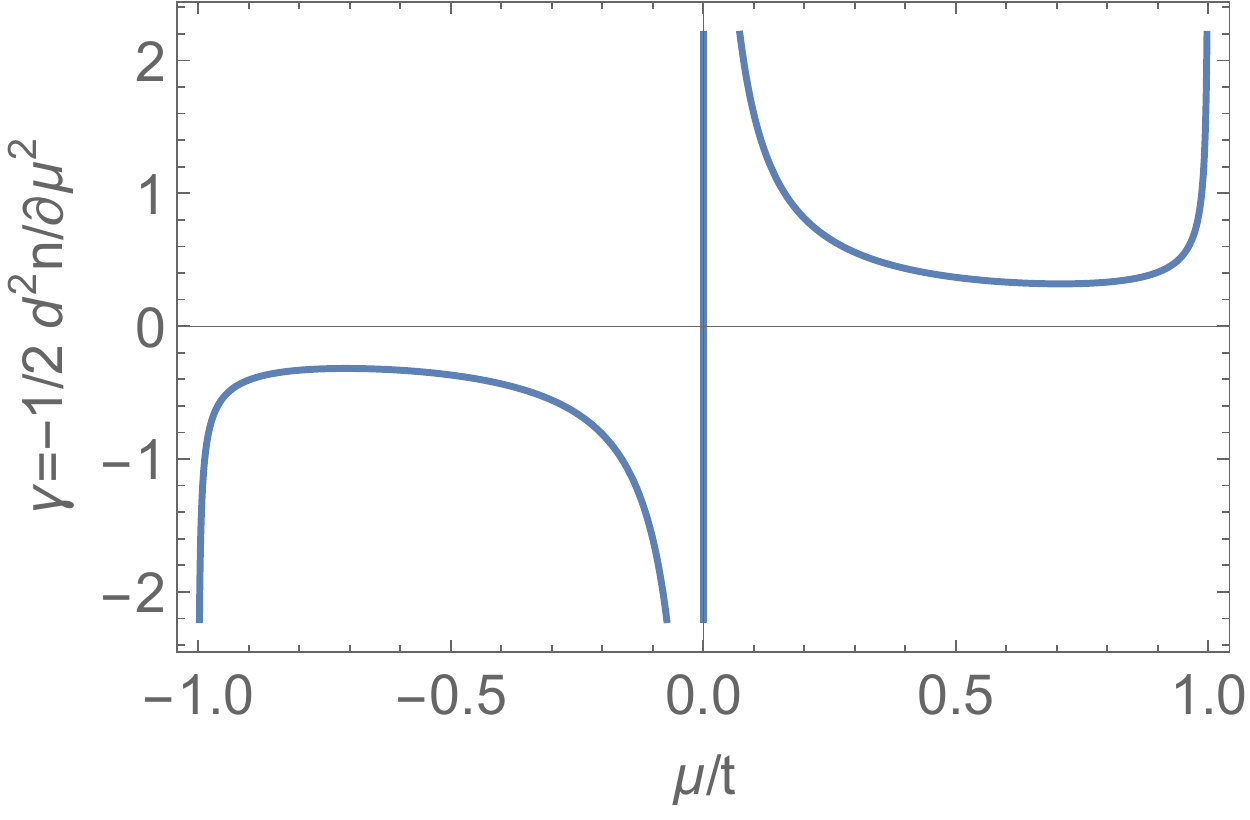}
\caption{Density $n$ as function of chemical potential $\mu$ according to Eq.~\eqref{eq:nofmu} (top left) and conversely chemical potential as function of filling. The first derivatives of $n$ determines the compressibility (bottom left) and the second derivative the bare cubic coefficient $\gamma$ in the free energy.}
\label{fig:nofmu}
\end{figure}

\begin{figure}[h]
\centering{}
\includegraphics[width=0.99\linewidth]{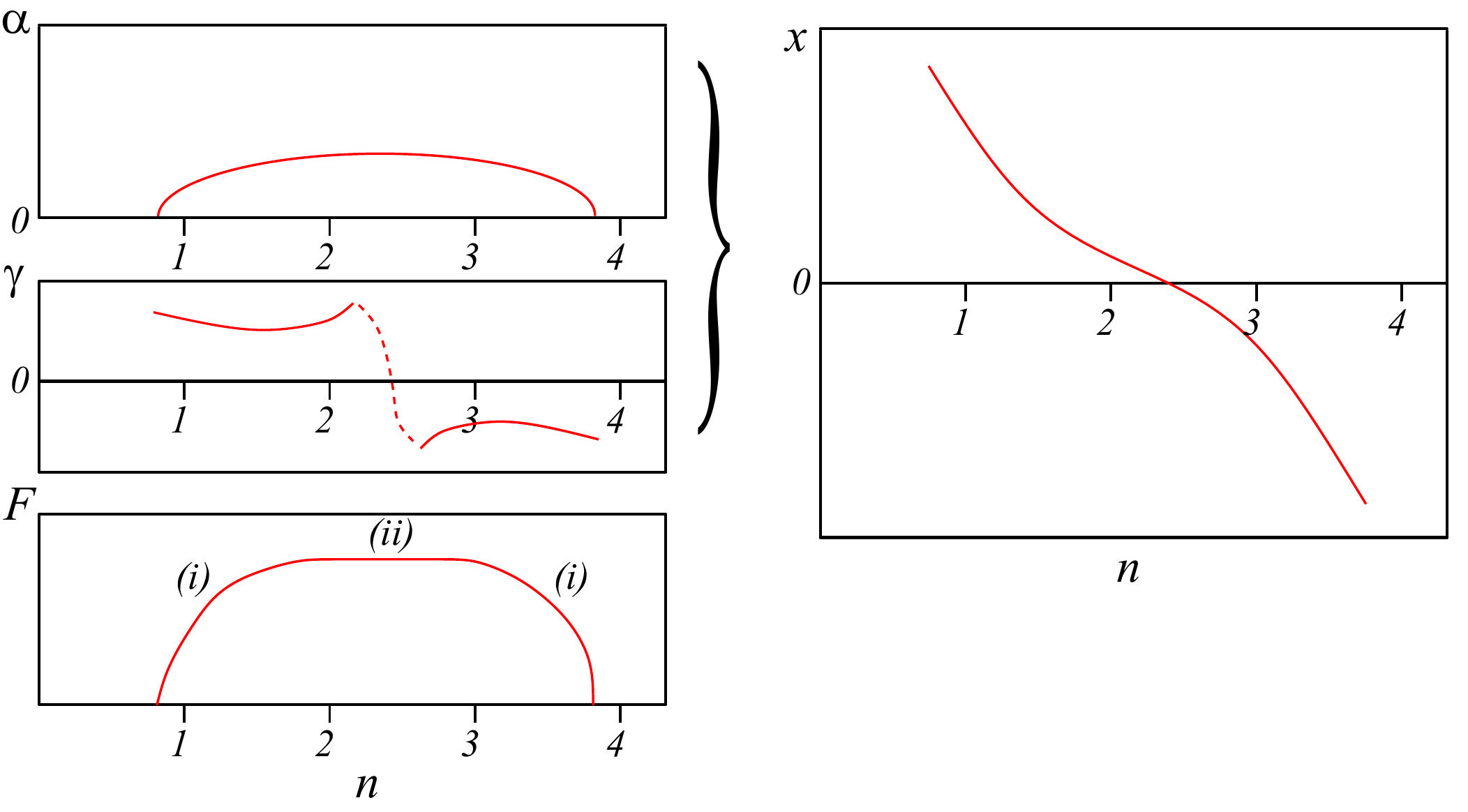}
\caption{Illustrative sketch of the behavior of $\alpha$, $\gamma$, $x$ as a function of filling $n$. The lowest picture illustrates the correspondence between the minima of the free energy and filling $n$. }
\label{pic_corr}
\end{figure}

We consider a generic dispersion around a van Hove point of the form $\xi({\bf k})=\epsilon({\bf k})-\mu$ with $\epsilon({\bf k})=k_x^2-k_y^2$ and $-t\leq \xi\leq t$. The filling for given chemical potential is given by
\begin{align}
n(\mu)&=\int d^2 k n_F(\xi({\bf k})) \notag\\
&= \frac{t\pi}{2}+\mu\text{arcoth}\frac{t}{\sqrt{t^2-\mu^2}}+t \arctan\frac{\mu}{\sqrt{t^2-\mu^2}}
\label{eq:nofmu}
\end{align}
with Fermi function $n_F$. This can also be expressed as
\begin{align}
n(\mu)=T\sum_{i\omega}\int d^2 k G_0(i\omega, {\bf k})\,,
\end{align}
where $G_0=(i\omega-\xi)^{-1}$ is the single-particle Green's function. Thus, the cubic coefficient $\gamma$ can be related to the single-particle density via
\begin{align}
\gamma&=-T\sum_{i\omega}\int d^2 k G_0^3(i\omega,{\bf k})\notag\\
&=-\frac{1}{2} \frac{d^2n}{d\mu^2}\,.
\end{align}
   This formula works at some distance from the van Hove point but not in its immediate vicinity because
   to derive it we interchanged integration and differentiation. This procedure is not valid at the
 van Hove point due to the singularity.
We show the density $n$ and its derivatives, including $\gamma$, in Fig.~\ref{fig:nofmu}. We see the expected sign change of $\gamma$  before and after the van Hove point. This behavior is generic around van Hove filling because the slope of the inverse compressibility $d\mu/dn$
 changes sign at the van Hove point.
 In Fig. \ref{pic_corr} we sketch how the Landau free energy parameters depend on electron filling.

\begin{figure}[h]
\centering{}
\includegraphics[width=0.5\linewidth]{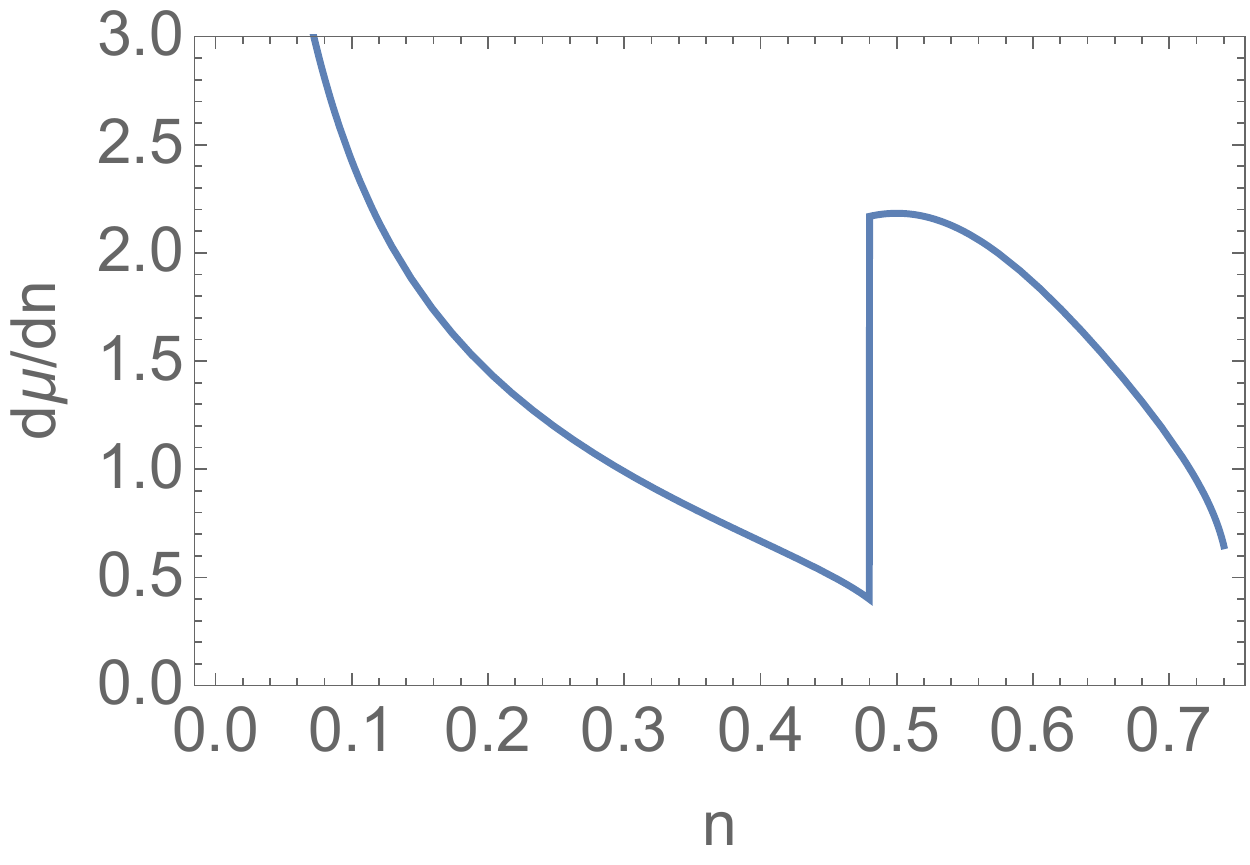}
\caption{Inverse compressibility as function of density $n$ assuming a first order transition that splits the van Hove peak according to $\epsilon_S({\bf k})=k_x^2-k_y^2\pm S$. We use $S=0.4$ for $n\geq 0.48$. }
\label{fig:sawtooth}
\end{figure}

In the main text we argue that, as the system approaches the van Hove energy at increasing doping, the van Hove peak is split in a first-order transition. As an example for the splitting, we consider an order parameter $S$, which splits the dispersion to $\epsilon_S({\bf k})=k_x^2-k_y^2\pm S$. This happens e.g., due to ferromagnetic order in both valleys around $n\approx 2$, that leads  to the 2-2 splitting.  We calculate the corresponding inverse compressibility in Fig.~\ref{fig:sawtooth}. An analogous behavior occurs at each of the transitions of the cascade, which reproduces to the seesaw behavior, observed in the measured compressibility (Fig. 1 of the main text).

\section*{Supplementary Discussion VI: Derivation of SU(3)-symmetric Landau functional from the microscopic 6-patch model}
\label{app:SU3}

\begin{figure}[h]
\centering{}
\includegraphics[width=0.6\linewidth]{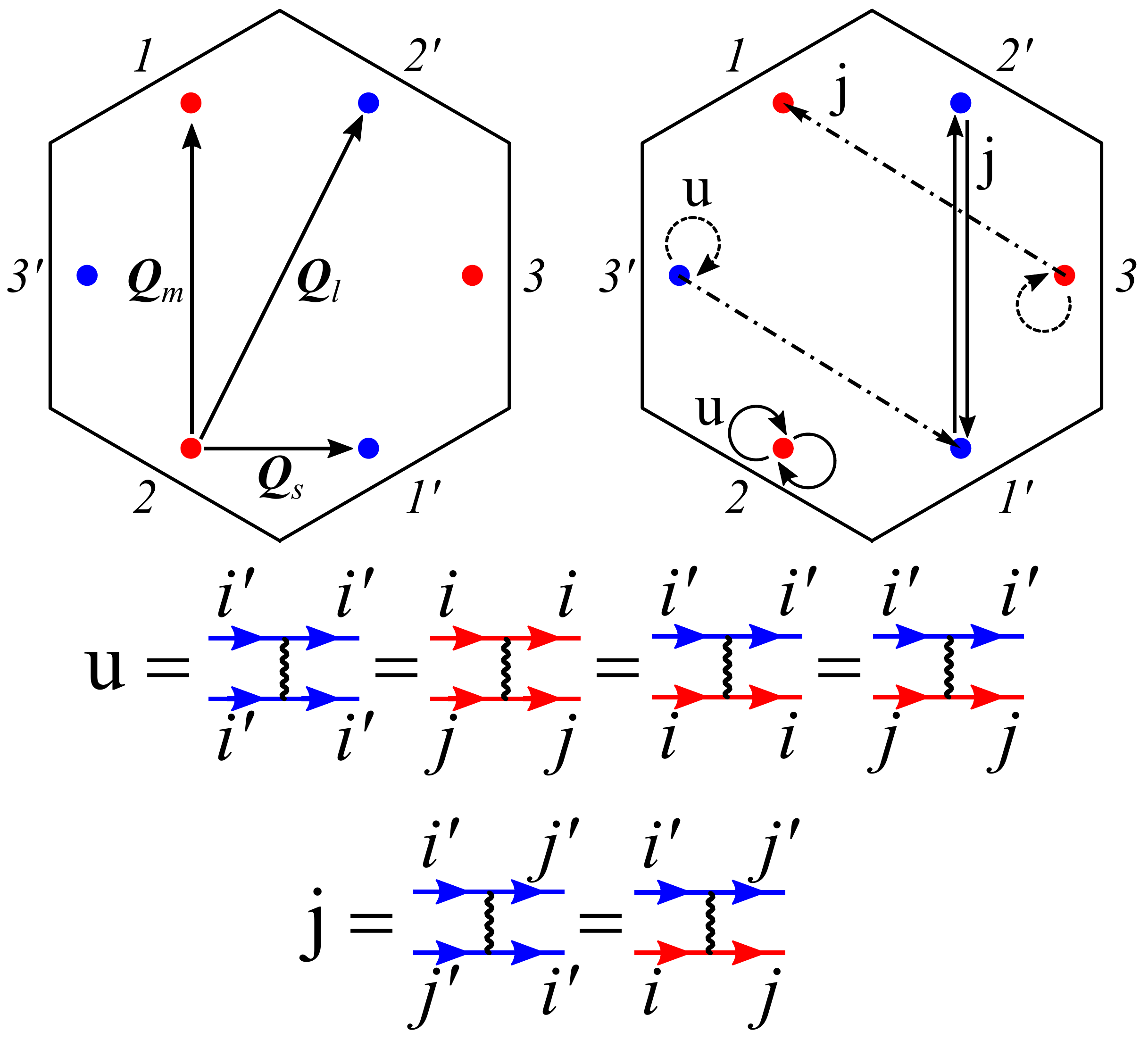}
\caption{Left: The sketch of a 6-patch model. Circles indicate the location of patches and color indicates valley composition. We assume that the blue valley is ``spinless''. Vectors $\bf Q_s, \bf Q_m, \bf Q_l$ connect patches in momentum space. Right: sketch of interactions between patch fermions. The couplings are shown below.}
\label{scat6p}
\end{figure}

Here we show how the SU(3) symmetry appears in the 6-patch model in the case of one valley hosting electrons with only one spin projection. The sketch of patch structure, momentum transfer vectors, and interactions is given in Fig. \ref{scat6p}. Density-density interactions between each
 patch are identical, exchange interaction is allowed if fermions do not change their valley in the process of scattering.  The coupling constants and polarization operators are the same as in the full 6-patch model with vH singularities of all 4 bands at the Fermi level. Every patch in the model we consider here is valley-polarized and one of the valleys is also spin-polarized. We label patches by  $i=1,2,3$ for valley $+$ and by $i'=1,2,3$ for valley $-$, spins by $s, s' = \uparrow, \downarrow$ and consider valley $-$ as spinless, i.e., having only one spin projection.
For the interaction we use the same model as in Ref. \cite{Chichinadze2020magnet}
 with density-density
 and exchange
 couplings $u$ and $j$.

 We introduce all possible order parameters involving fermions near vH points:
\begin{equation}
\begin{gathered}
\Delta_{Pom,+,i}^c = \langle f^{\dagger}_{+,s,i} \delta_{ss'} f_{+,s',i} \rangle, \\
\Delta_{Pom,+,i}^s = \langle f^{\dagger}_{+,s,i} \vec{\sigma}_{ss'} f_{+,s',i} \rangle, \\
\Delta_{Pom,-,i'} = \langle f^{\dagger}_{-,i'}  f_{-,i'} \rangle, \\
\Delta_{Q_s,+-,\up,i} = \langle f^{\dagger}_{+,\up,(i+2)}  f_{-,(i+1)'} \rangle, \\
\Delta_{Q_s,+-,\up,i'} = \langle f^{\dagger}_{+,\up,(i+1)}  f_{-,(i+2)'} \rangle, \\
\Delta_{Q_s,+-,\down,i} = \langle f^{\dagger}_{+,\down,(i+2)}  f_{-,(i+1)'} \rangle \\
\Delta_{Q_s,+-,\down,i'} = \langle f^{\dagger}_{+,\down,(i+1)}  f_{-,(i+2)'} \rangle \\
\Delta_{Q_m,+,i}^c = \langle f^{\dagger}_{+,s,i+2} \delta_{ss'} f_{+,s',i+1} \rangle, \\
\Delta_{Q_m,+,i}^s = \langle f^{\dagger}_{+,s,i+2} \vec{\sigma}_{ss'} f_{+,s',i+1} \rangle, \\
\Delta_{Q_m,-,i'} = \langle f^{\dagger}_{-,(i+2)'}  f_{-,(i+1)'} \rangle, \\
\Delta_{Q_l,+-,\up,i} = \langle f^{\dagger}_{+,\up,i}  f_{-,i'} \rangle, \\
\Delta_{Q_l,+-,\down,i} = \langle f^{\dagger}_{+,\down,i}  f_{-,i'} \rangle.
\end{gathered}
\label{delta_defs_6patch}
\end{equation}
There are also  conjugated parameters, which we omitted  for brevity.

We first consider intra-valley $Q=0$ channels.
There are six charge order parameters:
\begin{equation}
\begin{gathered}
\Gamma(Q=0)^c = \\  \left ( \Delta_{Pom,+,1}^c \; \Delta_{Pom,+,2}^c \Delta_{Pom,+,3}^c \; \Delta_{Pom,-,1'} \; \Delta_{Pom,-,2'} \; \Delta_{Pom,-,3'} \right). \notag
\end{gathered}
\end{equation}
 As usual,  we assume that bare order parameters  are infinitesimally small and dress them by the interactions
  in the ladder approximation.
  We obtain
  \begin{equation}
 \Gamma(Q=0)^c =  \Gamma(Q=0)^{c, (0)} + \Pi(0) \Lambda^c_{\vec Q =0} \Gamma(Q=0)^c,
  \end{equation}
  where $\Pi(0)$ is the polarization operator,  $\Gamma(Q=0)^{c, (0)}$ is bare (infinitesimal)
  order parameter, and
   $\Gamma(Q=0)^c$ is the dressed order parameter.
The matrix $\Lambda^c_{\vec Q =0}$  has the form
\begin{equation}
\Lambda^c_{\vec Q =0} =
\begin{pmatrix}
- u & j- 2u & j-2u & -u & -u & -u \\
j-2u & -u & j-2u & -u & -u & -u \\
j-2u & j-2u & -u & -u & -u & -u \\
-2u & -2u & -2u & 0 & j-u & j-u \\
-2u & -2u & -2u & j-u & 0 & j-u \\
-2u & -2u & -2u & j-u & j-u & 0
\end{pmatrix}.
\end{equation}
The largest eigenvalue of this
matrix is $\lambda_{Pom}^{c, s^{\pm}} = u + 2j$ and the corresponding eigenvector is $(-\frac{1}{2},-\frac{1}{2},-\frac{1}{2},1,1,1)$.
 It describes $s^{\pm}-$wave
 charge
  Pomeranchuk order, which is symmetric with respect to 3 vH points from the same valley and
  changes sign between valleys. This order leads to valley polarization.

A $Q=0$ spin order parameter can only be introduced for the spinfull valley, therefore
$$
\Gamma(Q=0)^s = \left ( \Delta_{Pom,+,1}^s \; \Delta_{Pom,+,2}^s \Delta_{Pom,+,3}^s  \right)
$$
with the coupling matrix
\begin{equation}
\Lambda^s_{\vec Q =0} =
\begin{pmatrix}
 u & j & j \\
j & u & j \\
j & j & u
\end{pmatrix}.
\end{equation}
The largest eigenvalue is again $u+2j$. The corresponding eigenvector is
 $(1,1,1)$, i.e., this order is again $s$-wave with respect to three vH points from the same valley.

Now consider intra-valley density wave orders.
 In general, such an order can be with any momenta connecting vH points.
 Intra-valley
  density-wave orders are with momenta $Q_m$ and inter-valley density-wave orders are with
 $Q_l$ and $Q_s$   (see Fig. \ref{scat6p}).

We begin with intra-valley orders.
As for $Q=0$, the
 order in the spin channel can be only introduced for the spinfull
  valley $+$.
The corresponding order parameters are
$$
\Gamma(Q_m)^s = \left ( \Delta_{Q_m,+,1}^s \; \Delta_{Q_m,+,2}^s \; \Delta_{Q_m,+,3}^s  \right)
$$
and the coupling matrix
is
\begin{equation}
\Lambda^s_{\vec Q_m} = u \mathbb{1}_{3\times 3}
\end{equation}
with three
eigenvalues  equal to $u$.
The charge channel is more interesting.
 Here
$$
\Gamma(Q_m)^c = \left ( \Delta_{Q_m,+,1}^c \; \Delta_{Q_m,+,2}^c \; \Delta_{Q_m,+,3}^c \; \bar{\Delta}_{Q_m,-,1'} \; \bar{\Delta}_{Q_m,-,2'} \; \bar{\Delta}_{Q_m,-,3'} \right),
$$
where $\bar{\Delta}$ is a conjugate of $\Delta$.
The matrix of the couplings is
\begin{equation}
\Lambda^c_{\vec Q_m} =
\begin{pmatrix}
u-2j & 0 & 0 & -j & 0 & 0 \\
0 & u-2j & 0 & 0 & -j & 0 \\
0 & 0 & u-2j & 0 & 0 & -j \\
-2j & 0 & 0 & u-j & 0 & 0 \\
0 & -2j & 0 & 0 & u-j & 0 \\
0 & 0 & -2j & 0 & 0 & u-j
\end{pmatrix}.
\end{equation}
The two eigenvalues of this matrix are $u$ and $u-3j$.

Now we proceed to inter-valley orders. Note, that for inter-valley orders  the
spin/charge dichotomy does
not
work anymore. Consider first the orders with momentum $Q_l$.
The order parameters  are
   \begin{equation}
   \begin{gathered}
\Gamma (Q_l) = \\ \left( \Delta_{Q_l,+-,\uparrow,1} \;  \Delta_{Q_l,+-,\uparrow,2} \; \Delta_{Q_l,+-,\uparrow,3} \; \Delta_{Q_l,+-,\downarrow,1} \;  \Delta_{Q_l,+-,\downarrow,2} \; \Delta_{Q_l,+-,\downarrow,3} \right).
\end{gathered}
\end{equation}
The coupling matrix is diagonal
\begin{equation}
\Lambda^s_{\vec Q_l} = u \mathbb{1}_{6\times 6}
\end{equation}
and has identical eigenvalues $u$.

We now consider
inter-valley orders with
 momentum  $Q_s$. The corresponding order parameters are
   \begin{equation}
   \begin{gathered}
\Gamma (Q_s) = \\ \left( \Delta_{Q_s,+-,\uparrow,1} \;  \Delta_{Q_s,+-,\uparrow,2} \; \Delta_{Q_s,+-,\uparrow,3} \; \Delta_{Q_s,+-,\uparrow,1'} \;  \Delta_{Q_s,+-,\uparrow,2'} \; \Delta_{Q_s,+-,\uparrow,3'} \right).
\end{gathered}
\end{equation}
 The coupling matrix is
\begin{equation}
\Lambda_{\vec Q_s} =
\begin{pmatrix}
u & 0 & 0 & j & 0 & 0 \\
0 & u & 0 & 0 & j & 0 \\
0 & 0 & u & 0 & 0 & j \\
j & 0 & 0 & u & 0 & 0 \\
0 & j & 0 & 0 & u & 0 \\
0 & 0 & j & 0 & 0 & u
\end{pmatrix}.
\end{equation}
The leading eigenvalue here is $u+j$.
It again corresponds to $s-$wave order, symmetric with respect to three vH points from the same valley
 (or, equivalently, symmetric with respect to three possible $Q_s$ between neighboring vH points from different valleys).
 Evaluating the products of the eigenvalues and the polarizations to obtain dimensionless couplings and
  comparing different channels, we find that s-wave  $Q=0$ channel and $s$-wave $Q_s$ channel are almost degenerate.
   This is the same type of degeneracy as in the model with spinfull fermions~\cite{Chichinadze2020magnet}.

  The outcome of this analysis is that there are 8 almost degenerate order parameters:
   one scalar intra-valley charge
   $Q=0$ order parameter, one 3-component intra-valley
   $Q=0$    vector    spin order parameter, and four  inter-valley order parameters
    with momenta $Q_s$, which one can treat as 4 scalars.
   These 8 order parameters form
   an
   adjoint representation of SU(3).

 The  matrix Green's function, symmetric with respect to 3 vH points from the same valley,
 is
a $3 \times 3$ matrix
 in
band space.
The Green's function of free fermions
 is diagonal and isotropic:
\begin{equation}
\hat{G}_0 =
\begin{pmatrix}
G & 0 & 0 \\
0 & G & 0 \\
0 & 0 & G
\end{pmatrix},
\end{equation}
where
$G^{-1}=i\omega_m-\xi_k$ with Matsubara frequency $\omega_m$ and fermion dispersion $\xi_k$,
We associate the bottom component
 with the spinless valley.
Once the order sets
in,
the Green's function gets modified.
 It is convenient to introduce
   the valley polarization order parameter $\phi$ via
\begin{align}
\Delta_{Pom,+}^c = \frac{1}{\sqrt{3}} \phi, \; \Delta_{Pom,-} = - \frac{2}{\sqrt{3}} \phi,
\end{align}
 and introduce  inter-valley
   order parameters
   $S_{A1}, S_{A2}, S_{B1}, S_{B2}$ related to the magnitudes of inter-valley order parameters
   (identical for 3 directions of vectors $\vec{Q}_s$)
    $\Delta_{Q_s,+-, \uparrow}, \Delta_{Q_s,+-, \downarrow}$ and their conjugated
   $\bar{\Delta}_{Q_s,+-, \uparrow}$ and
   $\bar{\Delta}_{Q_s,+-, \downarrow}$
 via
\begin{equation}
\begin{gathered}
S_{A2}-iS_{B2} = \Delta_{Q_s,+-, \uparrow}, \; S_{A1}- iS_{B1} = \Delta_{Q_s,+-, \downarrow}, \\
 S_{A2} + iS_{B2} = \bar{\Delta}_{Q_s,+-, \uparrow}, \; S_{A1} + i S_{B1} = \bar{\Delta}_{Q_s,+-, \downarrow},
\end{gathered}
\end{equation}
We label the three-component vector spin intra-valley order parameter  $\Delta_{Pom,+}^s$  as just
 $\vec S$.

In matrix notations we then have
$\hat{G} = \hat{G}_0 + \hat{\phi} + \hat{S} +  \hat{S}_{AB}$, where
\begin{equation}
\begin{gathered}
\hat{\phi} = \frac{1}{\sqrt{3}}
\begin{pmatrix}
\phi & 0 & 0 \\
0 & \phi & 0 \\
0 & 0 & -2\phi
\end{pmatrix}, \\
\hat{S} =
\begin{pmatrix}
S_z & S_x - i S_y & 0 \\
S_x + i S_y & -S_z & 0 \\
0 & 0 & 0
\end{pmatrix}, \\
\hat{S}_{AB} =
\begin{pmatrix}
0 & 0 & S_{A2} - i S_{B2} \\
0 & 0 & S_{A1} - i S_{B1} \\
S_{A2} + i S_{B2} & S_{A1} + i S_{B1} & 0
\end{pmatrix}.
\end{gathered}
\label{su3rep}
\end{equation}
Expressed via the standard form of Gell-Mann matrices,
 $\hat{\phi}$ corresponds to matrix $T^8$, $\hat{S}$ corresponds to $T^1, T^2, T^3$, and $\hat{S}_{AB}$ corresponds to $T^4, T^5, T^6, T^7$, where $T^i$ with $i=1...8$ are the eight generators of the group SU(3).

\section*{Supplementary Discussion VII: The ground state of the SU(3) model in terms of fermionic bilinears}

The ordered state in the SU(3) model
 can be straightforwardly expressed via fermionic bilinears.  For simplicity, we present the result for the case
 when inter-valley components $S_{A1}, S_{A2}$  and $ S_{B1}, S_{B2} $ are absent and the order is specified by $\phi$ and ${\bf S}$. The same free energy  as in (\ref{SU3free}), expressed in terms of $\phi$ and ${\bf S}$ is
  \beq
   F = \alpha (\phi^2 + \vec S \cdot \vec S) + \frac{2\gamma}{\sqrt{3}} \phi \left( \vec S \cdot \vec S - \frac{\phi^2}{3} \right) + \frac{\beta}{2} (\phi^2 + \vec S \cdot \vec S)^2
  \eeq
  We introduce a standard parameterization for magnitudes of order parameters: $|\vec S| = r \cos \theta, \; \phi = r \sin \theta$ and rewrite the free energy in the form
  \begin{equation}
  F = \alpha r^2 + \frac{2 \gamma}{3\sqrt{3}} r^3 \sin 3\theta + \frac{\beta}{2} r^4 .
  \end{equation}
  The ground state is reached when $\sin 3 \theta = -1$, i.e., for $\theta=\frac{\pi}{2},\frac{7\pi}{6},\frac{11\pi}{6} $.
  The ordered states,  which we earlier specified by $\hat{\Phi} = (\lambda, \lambda, -2\lambda)$
  (up to permutations)
   with $\lambda$ given by Eq. (\ref{lambdaSU3}), are
    expressed in terms of $\phi$ and $S$   as
    \begin{equation}
    \begin{gathered}
    \phi = r, |\vec S| = 0, \\
    \phi = -\frac{r}{2}, |\vec S| = -\frac{r\sqrt{3}}{2}, \\
    \phi = -\frac{r}{2}, |\vec S| = \frac{r\sqrt{3}}{2}.
    \end{gathered}
    \end{equation}
    The first state is a pure valley order, for which the two-fold degeneracy stems from the unbroken spin degeneracy of the spinfull valley. The other two states correspond to mixed spin-valley order. There, the degenerate levels necessarily belong to different valleys, however,
     spin directions remain degenerate.

  \section*{Supplementary Discussion VIII: Another scenario for the cascade of transitions}

Here we discuss another scenario for the cascade, in which
the component of the van Hove peak that
crosses the Fermi level,
no longer contributes to particle-hole order.
  This scenario is qualitatively similar to the one  put forward in Ref. \cite{Zondiner2020}, in which
  one of the bands gets fully filled (fully depleted) at each transition from the cascade
 and after that does not contribute to particle-hole order, and to the one in Ref. \cite{ali_2}, in which a flat band
 gets severely broadened after crossing the Fermi level.
  We do not assume full filling/full depletion or strong broadening, but still
  exercise here the idea that one of the bands effectively disappears after each transition, and the
    symmetry of the Landau free energy progressively reduces from SU(4) to SU(3) and then to SU(2). In this scenario the
        pattern of symmetry changes at the transitions from the cascade  is
\begin{align}
n &\approx 1: \; \mathrm{SU(4)} \rightarrow \mathrm{SU(3)}\times\mathrm{U(1)}\notag \\
n &\approx 2: \;  \mathrm{SU(3)} \rightarrow \mathrm{SU(2)}\times\mathrm{U(1)} \notag \\
n &\approx 3: \;   \mathrm{SU(2)} \rightarrow \mathrm{U(1)}.
\label{group_pattern}
\end{align}
 and the number of relevant bands that contribute to particle-hole orders changes from 4 to 3 at
 $|n| \approx 1$, from 3 to 2 at $|n| \approx 2$, and from 2 to 1 at $|n| \approx 3$.

The first transition and the manifold of the ordered states is  exactly the same as in the SU(4) model from the main text, and the splitting of the vH peak is 3-1, with one component crossing the Fermi level.  At the  second transition, one component of 3-fold degenerate vH peak crosses the Fermi level.  The manifold of the ordered states is
 the same as in the  2-2 phase of the SU(4) model.

 For the last transition, the relevant model contains either two spinless fermions from different valleys or one valley with both spin projections. In
 either
 case the system obeys SU(2) symmetry: in
 valley space in the former and in spin space in the latter. For the valley SU(2) case the only particle-hole order parameter is charge valley polarization.  Interaction for this order parameter is attractive, and gives rise to an instability that moves one vH peak component through the Fermi level and splits
 the
 doubly degenerate vH peak into 1-1. For the spin SU(2) case the only instability is the standard Stoner-like ferromagnetism that leads to the splitting of two levels.
 Finally, as $|n|$ comes close to 4, the last vH peak component moves through the Fermi level.

 Note that
 the
    first two transitions, near $|n|=1$ and $|n|=2$, are first order, the one near $|n|=3$ is second order, and the last crossing near $|n|=4$ is continuous in our present description, but may actually also involve a phase transition, as we argue in the next subsection.

  Within  this scenario, one
 can naturally explain the emergence of insulating states at integer fillings, but it is
  a priori
  unclear
  how the peak components which are assumed to be at different energies as they cross the Fermi level at different $n$,
  recombine
   back into a 4-fold degenerate strong vH peak once the order disappears at  $|n| \lesssim 4$.

 \section*{Supplementary Discussion IX: Instability of spinless fermions from a single valley }

     \begin{figure}[h]
\centering{}
\includegraphics[width=0.3\linewidth]{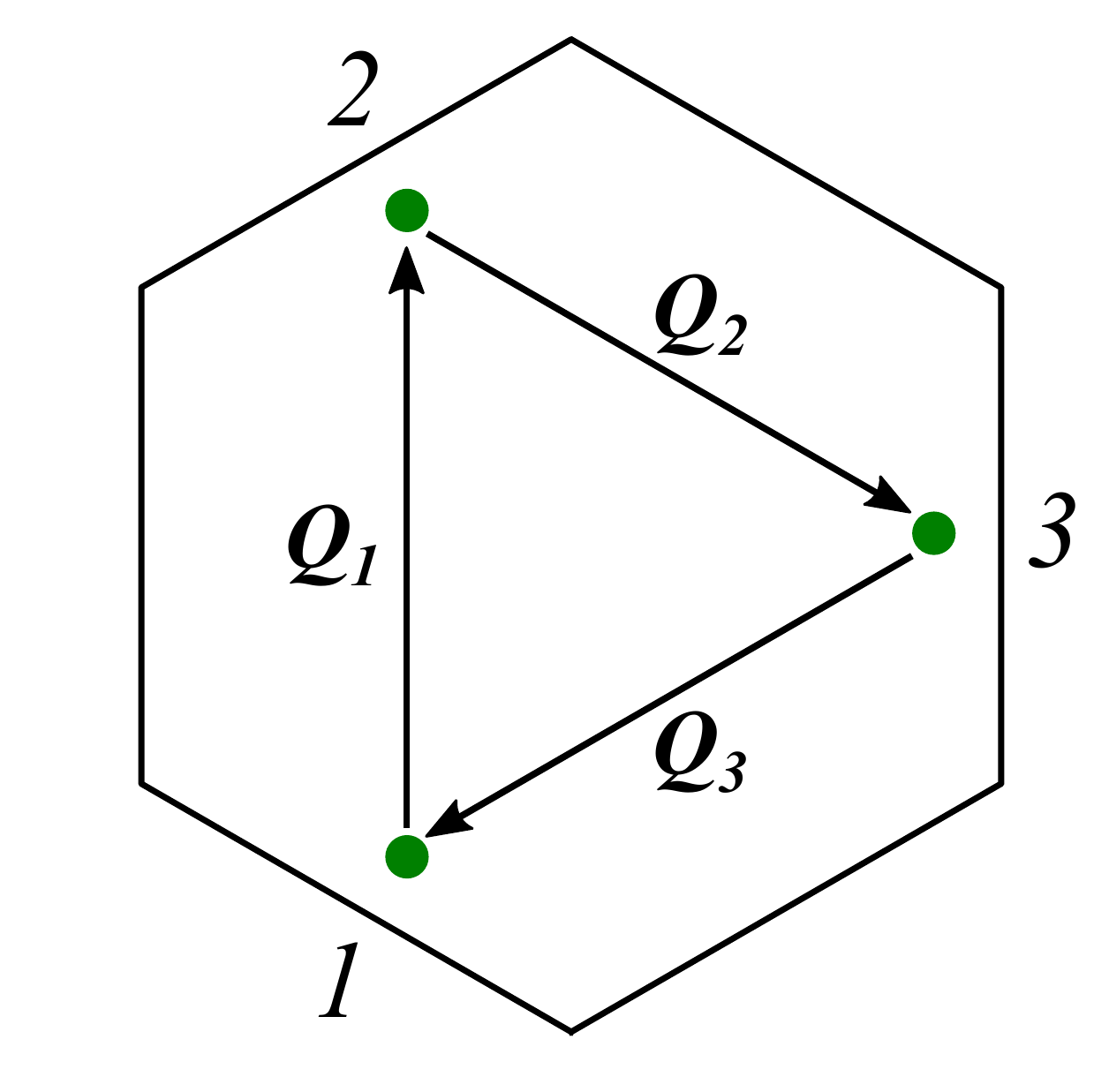}
\caption{The sketch of a patch model for a band with only one occupied valley. Green circles indicate the location of van Hove points. Three vectors indicate the directions of momentum transfer between each pair of patches. This is the momentum transfer of the density wave. }
\label{Qvec}
\end{figure}

Above we considered
vH points
that are
related by $C_3$ lattice rotational symmetry and assumed that magnitudes of order parameter are identical on every patch.  Here we relax this assumption and check if  other particle-hole orders are possible. This issue is most relevant
for
 the case of
 spinless fermions  near 3 vH points in only one valley
 i.e.,
 the case near $|n| \lesssim 4$
 when only one fermion species from a single valley remains.

 We label
 the fermions
 from the three
 vH points
 as $f_i$,  $i=1,2,3$, see Fig. \ref{Qvec}.
   Because fermions are spinless, only charge orders are possible. There are two
      potential orders: the one  with $Q=0$ and the one with $Q$ between van Hove points. Each order parameter has three
       components.
The order parameters are
\begin{equation}
\begin{gathered}
\Delta_{Pom, i} = \langle f^{\dagger}_i f_i \rangle, \\
\Delta_{\vec Q_i} = \langle f^{\dagger}_{i+1} f_i \rangle.
\end{gathered}
\end{equation}
Like in Supplementary Discussion VI we consider  density-density and exchange interactions ($u$ and $j$ terms, respectively).
 In the ladder approximation the two order parameters do not couple and can be considered independent of each other.
 For
 the
 three $Q=0$ order parameters the coupling matrix is
\begin{equation}
\Lambda_{Q=0} =
\begin{pmatrix}
0 & j-u & j-u \\
j - u& 0 & j-u \\
j-u& j-u & 0
\end{pmatrix}
\end{equation}
There are three eigenvalues: $\lambda^s = 2 (j -u)$, which corresponds to an $s-$wave order parameter
 with eigenvector $(1,1,1)$, and two-fold degenerate $\lambda^d = u - j$, with $d-$wave  eigenvectors  $(-1, 0, 1)$ and $(1/2,-1,1/2)$. For $u >j$, as expected on general grounds, $\lambda^d$ is positive (attractive) and $\lambda^s$ is negative (repulsive).
If the $s-$wave
 $Q=0$ order parameter
 is imposed,
 the vH peak crosses the Fermi level without inducing a particle-hole order.

For  finite $Q$ orders, the  coupling matrix is
\begin{equation}
\begin{gathered}
\Lambda_{Q} =
\begin{pmatrix}
u-j & 0 & 0 \\
0 & u-j & 0 \\
0 & 0 & u-j
\end{pmatrix}.
\end{gathered}
\end{equation}
  We see that the eigenfunctions for all three combinations of $\Delta_{Q_i}$  (one is $s-$wave and two are $d-$wave)
     are identical and the same as for the $d$-wave $Q=0$ order parameter.
The polarization operator for $Q=0$ is slightly larger than the one with   finite $Q$ (Ref. \cite{Chichinadze2020magnet}). Hence $d$-wave $Q=0$ order is the most likely one. Such an order splits the energies of the three vH peaks. So far, no
clear
evidence for such order has been reported,
see however, Ref.~\onlinecite{Wu2021}.
One option may be that such order oscillates between the two $d$-wave components at short spatial scales, and on average all three vH peaks move identically, like if there was no particle-hole order.

\section*{Supplementary Discussion X: Cascade of transitions in Bernal Bilayer Graphene and in Rhombohedral Trilayer Graphene in the 2-patch model}

In this section we discuss the extension of our analysis to cases of Bernal Bilayer Graphene (BBG) and Rhombohedral Trilayer Graphene (RTG). Band structure and Fermi surfaces of both systems are extremely similar, therefore, we consider them under one umbrella in this manuscript.

Our VH scenario can be successfully applied to the problem of cascade of electronic transitions in Bernal bilayer (BBG) and rhombohedral trilayer graphene (RTG).
Here we consider a
minimal 2-patch model to describe the cascade in BBG and RTG. We discuss an extension to a 6-patch model in the next section.

\begin{figure}[h]
\begin{minipage}[h]{0.99\linewidth}
\center{\includegraphics[width=0.99\linewidth]{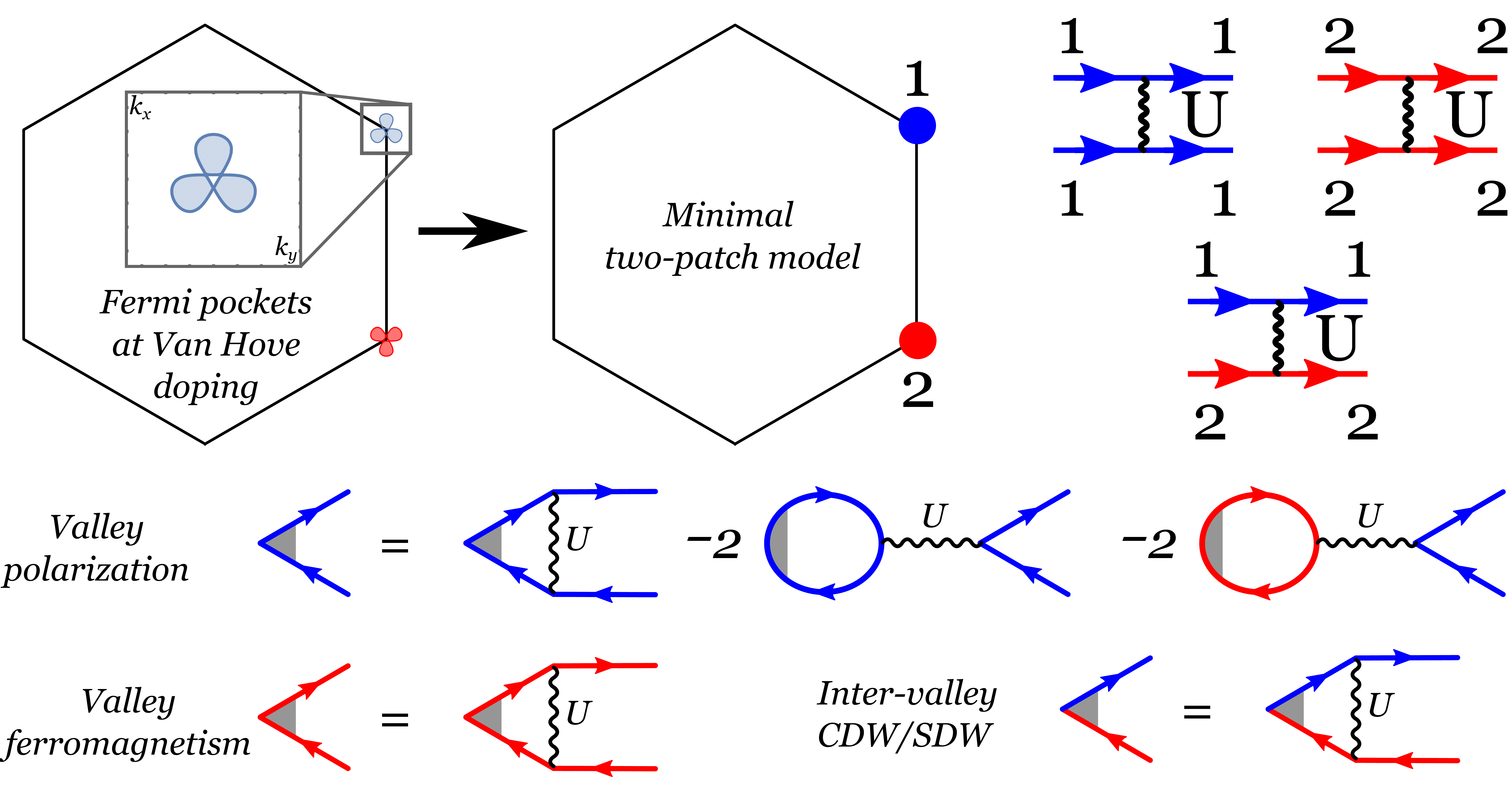} }
\end{minipage}
\centering{}\caption{
Top: sketch of Fermi surface
at vH doping
in Bernal bilayer graphene and rhombohedral trilayer graphene in a displacement field.
It can be approximated by a 2-patch model
with density-density interactions between patches 1 and 2.
Bottom: the system of coupled
equations for
 particle-hole instabilities in the 2-patch model.
}
\label{diag_2p}
\end{figure}

The two-patch model
describes
 fermions in
 the vicinity of K and K' points in the hexagonal Brillouin zone (BZ).
 This model is based on realistic tight-binding models for BBG \cite{McCann2006PRL,graphene_RMP,McCann_2013} and RTG \cite{Koshino2009ABCWarping,Zhang2010ABC} in the presence of a displacement field, which opens
a gap between the conduction and valence bands \cite{bilayer_bias}. We
assume that the chemical potential is near
vH doping close to charge neutrality.
In this case, the Fermi surface of both BBG and RTG either has the shape of three touching small Fermi pockets at $K$ and $K'$ points (6 Van Hove singularities total), or has one higher-order vH singularity (HOVHS) per valley located exactly at K (K'), see Fig. \ref{diag_2p}.
Because three vH  points per valley are located very close to each other in the BZ,
we assume that the couplings between them are the same and that the small difference in wave vectors connecting vH points within a valley can be neglected, i.e. possible wave vector transfers between vH points are approximately zero or $K-K'$. In this case, we can consider one patch per valley $K,K'$ that describes spin-degenerate fermionic states with dispersion $\varepsilon \simeq k_x^2-k_y^2$.

We model the interaction via equal intra- and inter-patch density-density couplings and
 neglect valley mixing terms.
This last assumption has been widely used for TBG.  Its validity for non-twisted BBG and RTG in the absence of a displacement field is not justified~\cite{aleiner,vafek_RG}, but we conjecture, following~\cite{zhiyu_22} that in the presence of a sizable displacement field exchange processes between the two valleys are small.

The interaction Hamitonian with density-density couplings reads
\begin{equation}
H_{int} = \sum_{i,j=1,2;  \sigma,\sigma' = \uparrow, \downarrow} \left( U f^{\dagger}_{i \sigma} f_{i \sigma} f^{\dagger}_{j \sigma'} f_{j \sigma'} \right),
\end{equation}
where $i$ is the patch (valley) index, and $\sigma, \sigma'$ are spin indices
(see Fig.~\ref{diag_2p}).
To study potential instabilities of the Fermi liquid within the RPA approach,
we write down the system of coupled equations for test vertices
in spin and charge channels for the two patches. Its diagrammatic representation in shown in Fig. \ref{diag_2p}.
In our
2-patch approximation,
the polarization bubble $\Pi(0)=\Pi(K-K')
$ so that $Q=0$ channels and density waves 
with $Q=K-K'$
are exactly degenerate. Moreover, those instabilities are described by exactly the same 15 fields $\phi_j$ and the same SU(4)-symmetric Landau free energy functional as in the case of TBG. This, in turn, yields the same physics of the cascade and the same resulting ground states as in the case of TBG.

\section*{Supplementary Discussion XI: Cascade of transitions in Bernal Bilayer Graphene and in Rhombohedral Trilayer Graphene in the 6-patch model}

\begin{figure}[h]
\begin{minipage}[h]{0.8\linewidth}
\center{\includegraphics[width=0.99\linewidth]{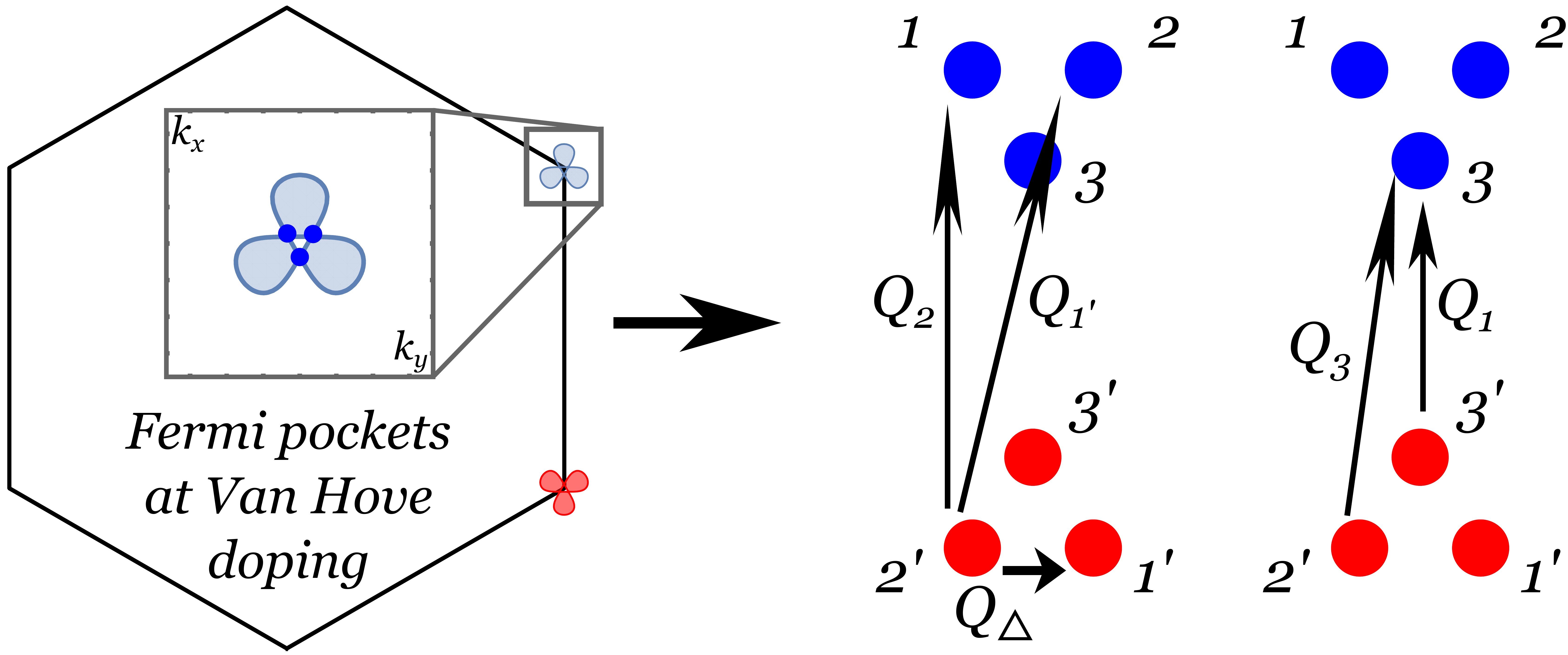} }
\end{minipage}
\centering{}\caption{
Sketch of the 6-patch model with one patch per vH point and different wave vectors connecting them.
Blue circles indicate positions of vH points. 
}
\label{6p_sketch}
\end{figure}

\begin{figure}[h]
\begin{minipage}[h]{0.5\linewidth}
\center{\includegraphics[width=0.8\linewidth]{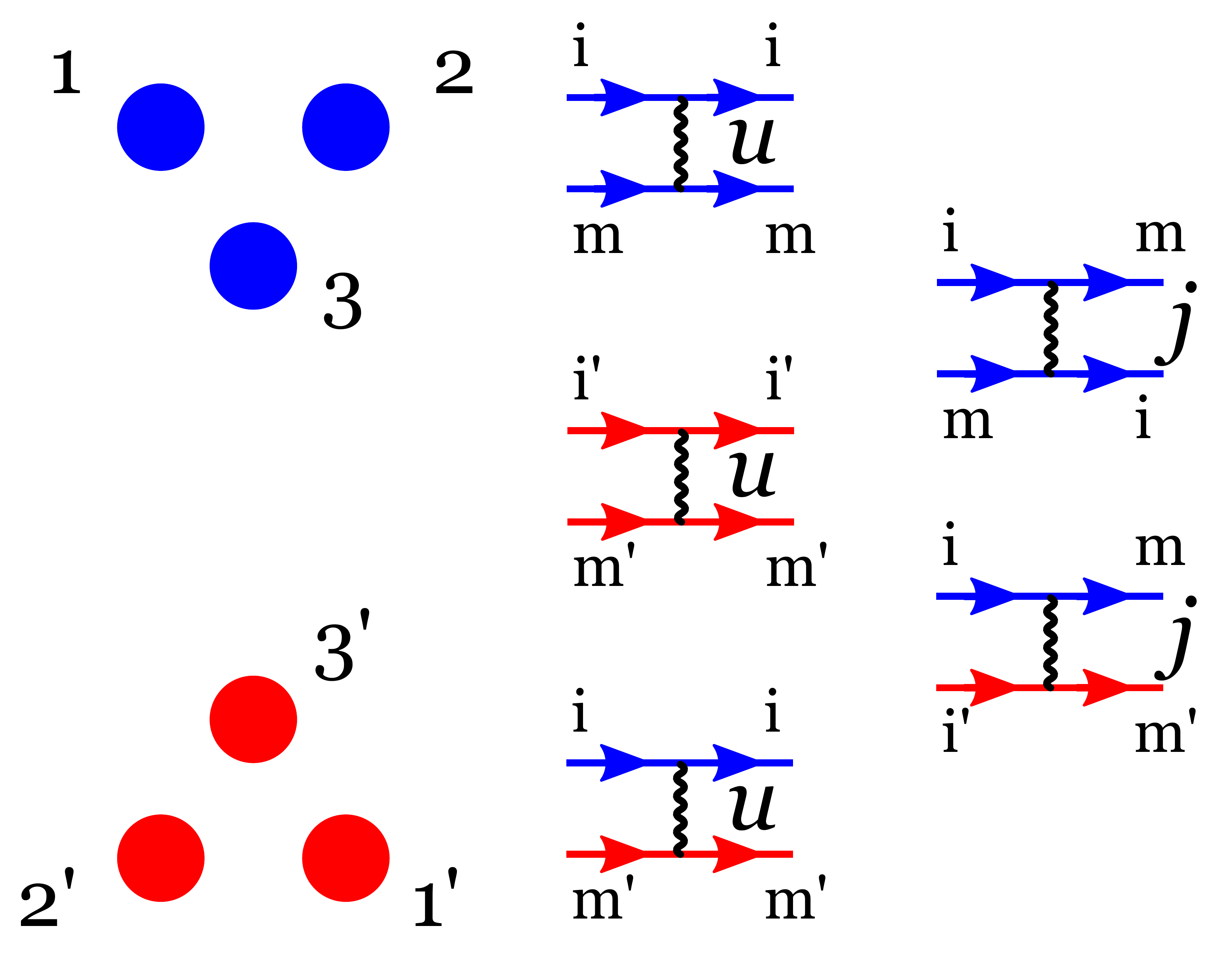} }
\end{minipage}
\centering{}\caption{
Density-density and exchange interactions between patch fermions, governed by the Hamiltonian (\ref{H_int}). }
\label{6p_sc}
\end{figure}

As a next step, we distinguish the states near the six vH points and model each of them via a hyperbolic dispersion relation valid in a patch around the vH point of the form $\varepsilon \simeq k_x^2-k_y^2$ and its rotations by $2\pi/3$, respectively.
The structure of patch model in momentum space is shown in Fig. \ref{6p_sketch}.
For the interaction, we can distinguish
five different magnitudes of momentum-transfer vectors. Some of them ($\vec Q_{\bigtriangleup}$) connect patches within a valley, the other $\vec Q_1, \vec Q_{1'}, \vec Q_2, \vec Q_3$ represent momentum transfers between different valleys.
Allowing for density-density couplings $u$ and exchange couplings $j$ between the different patches and neglecting valley mixing,
we obtain the
interaction Hamiltonian
\begin{equation}
\begin{gathered}
H_{int} = \sum_{i,m=1,2,3;  s,s' = \uparrow, \downarrow} \biggr[ u f^{\dagger}_{i,s,+} f_{i,s,+} f^{\dagger}_{m,s',+} f_{m,s',+} \\ + u f^{\dagger}_{i',s,-} f_{i',s,-} f^{\dagger}_{m',s',-} f_{m',s',-} + u f^{\dagger}_{i,s,+} f_{i,s,+} f^{\dagger}_{m',s',-} f_{m',s',-} \\
+ j f^{\dagger}_{i,s,+} f_{i+1,s,+} f^{\dagger}_{i+1,s',+} f_{i,s',+} + j f^{\dagger}_{i',s,-} f_{(i+1)',s,-} f^{\dagger}_{(i+1)',s',-} f_{i',s',-} \\
+  j f^{\dagger}_{i,s,+} f_{i+1,s,+} f^{\dagger}_{i',s',-} f_{(i+1)',s',-}
 \biggr],
\end{gathered}
\label{H_int}
\end{equation}
where $i,m$ are patch indices, $+,-$ label valley, and $s,s'$ are spin indices. The relevant scattering processes are shown in Fig. \ref{6p_sc}.

We classify the possible order parameters by the values of momentum transfer, valley composition, and if the instability is in charge or spin channel. In total, there are 144-1=143 possible components of particle-hole order parameters that can be cast into
a scalar (for charge order) or vector (for spin order) form like in the patch model treatment of twisted bilayer graphene \cite{Chichinadze2020magnet}.
 All these possible order parameters that involve fermions in the vicinity of vH points
are 
\begin{equation}
\begin{aligned}
&\Delta(0)^c_{i,+} = \langle f^{\dagger}_{s,i,+} \delta_{ss'} f_{s',i,+} \rangle, \; \Delta(0)^s_{i,+} = \langle f^{\dagger}_{s,i,+} \vec \sigma_{ss'} f_{s',i,+} \rangle, \\
&\Delta(0)^c_{i',-} = \langle f^{\dagger}_{s,i',-} \delta_{ss'} f_{s',i',-} \rangle, \; \Delta(0)^s_{i',-} = \langle f^{\dagger}_{s,i',-} \vec \sigma_{ss'} f_{s',i',-} \rangle, \\
&\Delta(\vec Q_{\bigtriangleup})^c_{i,+} = \langle f^{\dagger}_{s,i+2,+} \delta_{ss'} f_{s',i+1,+} \rangle, \\
&\Delta(\vec Q_{\bigtriangleup})^s_{i,+} = \langle f^{\dagger}_{s,i+2,+} \vec \sigma_{ss'} f_{s',i+1,+} \rangle, \\
&\Delta(\vec Q_{\bigtriangleup})^c_{i',-} = \langle f^{\dagger}_{s,i'+2,-} \delta_{ss'} f_{s',i'+1,-} \rangle,\\
&\Delta(\vec Q_{\bigtriangleup})^s_{i',-} = \langle f^{\dagger}_{s,i'+2,-} \vec \sigma_{ss'} f_{s',i'+1,-} \rangle, \\
&\Delta(\vec Q_{1})^c = \langle f^{\dagger}_{s,3,+} \delta_{ss'} f_{s',3',-} \rangle, \; \Delta(\vec Q_{1})^s = \langle f^{\dagger}_{s,3,+} \vec \sigma_{ss'} f_{s',3',-} \rangle, \\
&\Delta(\vec Q_{1'})^c_{i=1,2} = \langle f^{\dagger}_{s,i,+} \delta_{ss'} f_{s',i',-} \rangle, \; \Delta(\vec Q_{1'})^s_{i=1,2} = \langle f^{\dagger}_{s,i,+} \vec \sigma_{ss'} f_{s',i',-} \rangle, \\
&\Delta(\vec Q_{2})^c_1 = \langle f^{\dagger}_{s,1,+} \delta_{ss'} f_{s',2',-} \rangle, \; \Delta(\vec Q_{2})^s_1 = \langle f^{\dagger}_{s,1,+} \vec \sigma_{ss'} f_{s',2',-} \rangle, \\
&\Delta(\vec Q_{2})^c_2 = \langle f^{\dagger}_{s,2,+} \delta_{ss'} f_{s',1',-} \rangle, \; \Delta(\vec Q_{2})^s_2 = \langle f^{\dagger}_{s,2,+} \vec \sigma_{ss'} f_{s',1',-} \rangle, \\
&\Delta(\vec Q_{3})^c_{1} = \langle f^{\dagger}_{s,3,+} \delta_{ss'} f_{s',1',-} \rangle, \; \Delta(\vec Q_{3})^s_{1} = \langle f^{\dagger}_{s,3,+} \vec \sigma_{ss'} f_{s',1',-} \rangle, \\
&\Delta(\vec Q_{3})^c_{2} = \langle f^{\dagger}_{s,3,+} \delta_{ss'} f_{s',2',-} \rangle, \; \Delta(\vec Q_{3})^s_{2} = \langle f^{\dagger}_{s,3,+} \vec \sigma_{ss'} f_{s',2',-} \rangle, \\
&\Delta(\vec Q_{3})^c_{1'} = \langle f^{\dagger}_{s,1,+} \delta_{ss'} f_{s',3',-} \rangle, \; \Delta(\vec Q_{3})^s_{1'} = \langle f^{\dagger}_{s,1,+} \vec \sigma_{ss'} f_{s',3',-} \rangle, \\
&\Delta(\vec Q_{3})^c_{2'} = \langle f^{\dagger}_{s,2,+} \delta_{ss'} f_{s',3',-} \rangle, \; \Delta(\vec Q_{3})^s_{2'} = \langle f^{\dagger}_{s,2,+} \vec \sigma_{ss'} f_{s',3',-} \rangle,
\end{aligned}
\end{equation}
where $+,-$ labels the two valleys, $s,s'$ labels spin, $i=1,2,3$
are the patch numbers unless
specified otherwise, and $\vec \sigma$ is the vector of Pauli matrices.
For brevity, we did not list complex conjugates of the order parameters.

To determine possible Fermi liquid instabilities, we introduce infinitesimally small bare order parameters $\Gamma(Q)^{c,s,(0)}_{inter/intra}$ and consider their dressing by interactions in the ladder (RPA) approximation. In this approximation the dressed order parameter is given by
$$
\Gamma(Q)^{c,s}_{inter/intra} = \Gamma(Q)^{c,s,(0)}_{inter/intra} + \Pi (Q) \Lambda(Q)^{c,s}_{inter/intra} \Gamma(Q)^{c,s}_{inter/intra},
$$
where the matrix $\Lambda(Q)^{c,s}_{inter/intra}$
contains the couplings between
test vertices.
We start by considering the $Q=0$ channels. For the $Q=0$ charge channel
$$
\Gamma(0)^c = \left( \Delta(0)^c_{1,+}, \Delta(0)^c_{2,+}, \Delta(0)^c_{3,+}, \Delta(0)^c_{1',-}, \Delta(0)^c_{2',-}, \Delta(0)^c_{3',-} \right)
$$
and the coupling matrix reads
\begin{equation}
\Lambda(0)^c =
\begin{pmatrix}
-u & j-2u & j-2u & -2u & -2u & -2u \\
j-2u & -u & j-2u & -2u & -2u & -2u \\
j-2u & j-2u & -u & -2u & -2u & -2u \\
 -2u & -2u & -2u & -u & j-2u & j-2u \\
  -2u & -2u & -2u & j-2u & -u & j-2u \\
   -2u & -2u & -2u & j-2u & j-2u & -u
\end{pmatrix}.
\end{equation}
Its largest eigenvalue $u+2j$ corresponds to a valley charge order with an $s^{\pm}$ form factor $(1,1,1,-1,-1,-1)$.
The test vertex for a
$Q=0$ instability in the spin channel is
$$
\Gamma(0)^s = \left( \Delta(0)^s_{1,+}, \Delta(0)^s_{2,+}, \Delta(0)^s_{3,+}, \Delta(0)^s_{1',-}, \Delta(0)^s_{2',-}, \Delta(0)^s_{3',-} \right)
$$
and
the coupling matrix is
\begin{equation}
\Lambda(0)^s =
\begin{pmatrix}
u & j & j & 0 & 0 & 0 \\
j & u & j & 0 & 0 & 0 \\
j & j & u & 0 & 0 & 0 \\
 0 & 0 & 0 & u & j & j \\
  0 & 0 & 0 & j & u & j \\
   0 & 0 & 0 & j & j & u
\end{pmatrix}.
\end{equation}
The largest eigenvalue is again $u+2j$. It corresponds to two eigenvectors $(1,1,1,0,0,0)$ and $(0,0,0,1,1,1)$ that describe two independent valley ferromagnets with arbitrary orientation of magnetization vectors.
 Thus, the instabilities towards valley charge order and valley ferromagnetism are degenerate.

In contrast to the 2-patch model, there are also intra-valley spin (SDW) and charge (CDW) density waves with momentum transfer $\vec Q_{\bigtriangleup}$ in the 6-patch model. 
The test vertex for the CDW reads
\begin{align}
&\Gamma(\vec Q_{\bigtriangleup})^c = \\
& \left( \Delta(\vec Q_{\bigtriangleup})^c_{1,+}, \Delta(\vec Q_{\bigtriangleup})^c_{2,+}, \Delta(\vec Q_{\bigtriangleup})^c_{3,+}, \Delta(\vec Q_{\bigtriangleup})^c_{1',-}, \Delta(\vec Q_{\bigtriangleup})^c_{2',-}, \Delta(\vec Q_{\bigtriangleup})^c_{3',-} \right) \nonumber
\end{align}
and the coupling matrix is given by
\begin{equation}
\Lambda(\vec Q_{\bigtriangleup})^c =
\begin{pmatrix}
u-2j & 0 & 0 & -2j & 0 & 0 \\
0 & u-2j & 0 & 0 & -2j & 0 \\
0 & 0 & u-2j & 0 & 0 & -2j \\
 -2j & 0 & 0 & u-2j & 0 & 0 \\
  0 & -2j & 0 & 0 & u-2j & 0 \\
   0 & 0 & -2j & 0 & 0 & u-2j
\end{pmatrix}.
\end{equation}
The maximal eigenvalues of this coupling matrix is $u$. For intra-valley SDW with
\begin{align}
&\Gamma(\vec Q_{\bigtriangleup})^s =\\
& \left( \Delta(\vec Q_{\bigtriangleup})^s_{1,+}, \Delta(\vec Q_{\bigtriangleup})^s_{2,+}, \Delta(\vec Q_{\bigtriangleup})^s_{3,+}, \Delta(\vec Q_{\bigtriangleup})^s_{1',-}, \Delta(\vec Q_{\bigtriangleup})^s_{2',-}, \Delta(\vec Q_{\bigtriangleup})^s_{3',-} \right) \nonumber
\end{align}
the coupling matrix is diagonal
$$
\Lambda(\vec Q_{\bigtriangleup})^s = u \mathbb{1}_{6\times6},
$$
where $ \mathbb{1}_{6\times6}$ is a $6\times 6$ diagonal matrix in patch space.
Hence, for intra-valley channels, instabilities towards charge and spin density waves are 
degenerate within RPA.

We now move to inter-valley channels.
 The couplings in charge and spin inter-valley density wave channels are degenerate.  The  density-wave order parameter with momentum $Q_1$ and $\Gamma(\vec Q_{1})^{c,s} = \Delta(\vec Q_{1})^{c,s}$ couples only to
 itself. Therefore, the only eigenvalue is $u$. For $Q_{1'}$, $\Gamma(\vec Q_{1'})^{c,s} = \left( \Delta(\vec Q_{1'})^{c,s}_1 , \Delta(\vec Q_{1'})^{c,s}_2 \right) $ has two components and the coupling matrix is diagonal
\begin{equation}
\Lambda(\vec{Q}_{1'})^{c,s} = u \mathbb{1}_{2\times2}.
\end{equation}
 For the order parameter with momentum $Q_2$, with
$$
\Gamma(\vec Q_{2})^{c,s} = \left( \Delta(\vec Q_{2})^{c,s}_{1} , \Delta(\vec Q_{2})^{c,s}_{2} \right),
$$
 the coupling matrix
 reads
\begin{equation}
\Lambda(\vec{Q}_{2})^{c,s} =
\begin{pmatrix}
u & j \\
j & u
\end{pmatrix}
\end{equation}
and the largest eigenvalue is $u+j$.
The coupling matrix for $\vec Q_3$ with
$$
\Gamma(\vec Q_{3})^{c,s}_1 = \left( \Delta(\vec Q_{2})^{c,s}_{1} , \Delta(\vec Q_{2})^{c,s}_{1'} \right)
$$
and
$$
\Gamma(\vec Q_{3})^{c,s}_2 = \left( \Delta(\vec Q_{2})^{c,s}_{2} , \Delta(\vec Q_{2})^{c,s}_{2'} \right)
$$
is identical to $\Lambda(\vec{Q}_{2})^{c,s}$. Hence, it yields the same leading eigenvalue.

In order to find the leading instability we need to know the values of polarization operators.
We find that
the two largest polarization operators are $\Pi(0)
>\Pi(Q_2)$ with nearly equal values. The
other
  $\Pi (Q_{i \neq 2})$
  are somewhat smaller in magnitude.
Neglecting the other finite $Q$ channels,
we obtain that
the
leading
instabilities are degenerate $Q=0$ valley ferromagnetism and charge valley order,
or degenerate inter-valley spin/charge density waves with momentum transfer of magnitude $Q_2$.
The instability towards $Q=0$ orders occurs when
\begin{equation}
1-\Pi(0) (u +2j) =0
\end{equation}
and for $Q=Q_2$ when
\begin{equation}
1-\Pi(Q_2) (u +j)=0.
\end{equation}
  For $j>0$, as we assumed to hold,
  the leading instabilities are with $Q=0$, i.e. two independent intra-valley ferromagnetic instabilities (every patch develops ferromagnetism independent from the other one) and a valley charge instability. This valley charge order results in
a different population of the two valleys.
For valley charge order the order parameter
is of the form $\phi = \sum_{i=1}^3 \av{f^{\dagger}_{s,i,+} \delta_{ss'} f_{s',i,+}} - \av{f^{\dagger}_{s,i',-} \delta_{ss'} f_{s',i',-}}$ and for the two valley ferromagnets
the order parameters are $\vec S_{1} = \sum_{i=1}^3 \av{f^{\dagger}_{s,i,+} \vec{\sigma}_{ss'}  f_{s',i,+}}$ and $\vec S_{2} =  \sum_{i=1}^3 \av{f^{\dagger}_{s,i',-} \vec{\sigma}_{ss'}  f_{s',i',-}}$.  Thus, the leading instabilities are described by the order parameter matrix $\hat{\Phi}_7$ from \eqref{phi_matr_7_8}.
As we showed above, this leads to the same cascade of transitions as in the SU(4) symmetric case.
We note that if $\Pi(Q_2)$ becomes larger than $\Pi(0)$ (and $j$ remains small), spin and charge density waves with wave vector $Q_2$ are the leading instability. In this case, the symmetry between the three pockets around $K,K'$ is broken together with threefold rotation symmetry.

\end{document}